\documentclass[12pt]{article}

\usepackage{amsmath}
\usepackage{caption}
\usepackage{subfig}
\usepackage{graphicx} % figuras
\usepackage{amsmath}
\usepackage{graphics} % figuras
\usepackage{anysize}
\usepackage{multirow}%%
\usepackage{wrapfig}
\usepackage{float}
\usepackage[usenames]{color}
\usepackage[hidelinks]{hyperref}

\usepackage{afterpage}

\usepackage{scicite}
\usepackage{times}

\newenvironment{sciabstract}{%
\begin{quote} \bf}
{\end{quote}}

\newcounter{lastnote}

\title{Molecular Identification with Atomic Force Microscopy and Conditional Generative Adversarial Networks} 

\author
{Jaime Carracedo-Cosme$^{1,2}$, Rub\'en P\'erez$^{2,3,\ast}$\\
\\
\normalsize{$^{1}$Quasar Science Resources S.L.,}\\ 
\normalsize{Camino de las Ceudas 2, E-28232 Las Rozas de Madrid, Spain}\\
\normalsize{$^{2}$Departamento de F\'isica Te\'orica de la Materia Condensada,}\\ \normalsize{Universidad Aut\'onoma de Madrid, E-28049 Spain}\\
\normalsize{$^{3}$Condensed Matter Physics Center (IFIMAC),}\\ 
\normalsize{Universidad Aut\'onoma de Madrid, E-28049 Madrid, Spain}\\
\\
\normalsize{$^\ast$ ruben.perez@uam.es}
}

\date{May 1, 2022}

\usepackage[acronym]{glossaries}
\newacronym{AFM}{AFM}{Atomic Force Microscopy}
\newacronym{NCAFM}{NC--AFM}{non--contact Atomic Force Microscopy}
\newacronym{HRAFM}{HR--AFM}{High Resolution Atomic Force Microscopy}

\newacronym{VAE}{VAE}{Variational Autoencoder}
\newacronym{CNN}{CNN}{Convolutional Neural Network}
\newacronym{MLP}{MLP}{Multilayer Perceptron}
\newacronym{RNN}{RNN}{Recurrent Neural Network}
\newacronym{NLP}{NLP}{Natural Languaje Procesing}
\newacronym{M-RNN}{M-RNNs}{Multimodal Recurrent Neural Networks}
\newacronym{LSTM}{LSTM}{Long Short--Term Memory}
\newacronym{DL}{DL}{Deep Learning}
\newacronym{GRU}{GRU}{Gated Recurrent Unit}
\newacronym{NN}{NN}{Neural Networks}
\newacronym{GUI}{GUI}{Graphical User Interface}
\newacronym{XOR}{XOR}{exclusive--or}
\newacronym{MARNN}{AM--RNN}{Attribute Multimodal Recurrent Neural Network}
\newacronym{SELFIES}{SELFIES}{Self-Referencing Embedded Strings}
\newacronym{GPU}{GPUs}{Graphical Processing Units}

\newacronym{AvgPool}{AvgPool}{Average Pool Layer}
\newacronym{MaxPool}{MaxPool}{Maximum Pool Layer}

\newacronym{lrelu}{LReLU}{Leaky ReLU}
\newacronym{relu}{ReLU}{Rectified Linear Unit Activation Function}
\newacronym{elu}{ELU}{Exponential Linear Unit}
\newacronym{MSE}{MSE}{Mean Squared Error}
\newacronym{MAE}{MAE}{Mean Absolute Error}
\newacronym{RMSE}{RMSE}{Root Mean Square Error}

\newacronym{RMSProp}{RMSProp}{Resilent Mean Square Propagation}
\newacronym{SGD}{SGD}{Stochastic Gradient Descent}
\newacronym{Adagrad}{Adagrad}{Adaptive Gradient Descent}
\newacronym{Nadam}{Nadam}{Nesterov Adaptive Moment Estimator}
\newacronym{Adam}{Adam}{Adaptive Moment Estimator}
\newacronym{CGAN}{CGAN}{Conditional Generative Adversarial Network}
\newacronym{GAN}{GAN}{Generative Adversarial Network}

\newacronym{lr}{lr}{learning rate}
\newacronym{AFMD}{QUAM--AFM}{Quasar Science Resources -- Autonomous University of Madrid Atomic Force Microscopy Image Dataset}
\newacronym{FC}{FC}{Fully connected}
\newacronym{IDG}{IDG}{Image Data Generator}
\newacronym{PPM}{PPM}{Probe Particle Model}
\newacronym{FDBM}{FDBM}{Full Density Based Model}
\newacronym{IUPAC}{IUPAC}{International Union of Pure and Applied Chemistry}
\newacronym{softmax}{Softmax}{Soft Approximation of Max}
\newacronym{AI}{AI}{Artificial Intelligence}

\newacronym{dAFM}{dAFM}{Dynamic Atomic Force Microscopy}
\newacronym{AM-AFM}{AM--AFM}{Amplitude Modulation Atomic Force Microscopy}
\newacronym{CH}{CH}{Constant Height}
\newacronym{DFT}{DFT}{Density Functional Theory}
\newacronym{ES}{ES}{Electrostatic}
\newacronym{SR}{SR}{Short--Range}
\newacronym{vdW}{vdW}{van der Waals}
\newacronym{PTCDA}{PTCDA}{perylene-tetracarboxylic-dianhydride}
\newacronym{FM-AFM}{FM--AFM}{Frequency Modulation Atomic Force Microscopy}
\newacronym{KPFM}{KPFM}{Kelvin Probe Force Microscopy}

\newacronym{LJ}{LJ}{Lennard--Jones}
\newacronym{NC}{NC}{Non--Contact}
\newacronym{NTCDI}{NTCDI}{Naphthalenetetracarboxylic diimide}
\newacronym{PAW}{PAW}{Projector--Augmented--Wave}
\newacronym{PES}{PES}{Potential Energy Surface}
\newacronym{SAM}{SAM}{Self--Assembled Monolayer}
\newacronym{SPM}{SPM}{Scanning Probe Microscopy}
\newacronym{STM}{STM}{Scanning Tunneling Microscopy}
\newacronym{STS}{STS}{Scanning Tunneling Spectroscopy}
\newacronym{SWCNT}{SWCNT}{Single Wall Carbon Nanotubes}
\newacronym{UHV}{UHV}{Ultra--High Vacuum}
\newacronym{LT}{LT}{Low Temperature}
\newacronym{VASP}{VASP}{Vienna Ab initio Simulation Package}
\newacronym{XC}{XC}{Exchange and Correlation}
\newacronym{HR}{HR}{High--Resolution}
\newacronym{TH}{TH}{Tersoff--Hamann}
\newacronym{OpenMX}{OpenMX}{Open source package for Material eXplorer}
\newacronym{C60}{C$_{60}$}{Buckminsterfullerene}
\newacronym{CVD}{CVD}{Chemical Vapor Deposition}
\newacronym{bfA}{bfA}{Breitfussin A}
\usepackage{gensymb}

\newacronym{BLEU}{BLEU}{Bilingual Evaluation Understudy}
\newacronym{M-RNN-AT}{M--RNN$_{A}$}{Multimodal Recurrent Neural Network predicting attributes}
\newacronym{mDBPc}{mDBPc}{meso--Dibenzoporphycene}

\newacronym{RES}{RES}{Red Espa\~n{}ola de Supercomputaci\'{o}n}

\usepackage{color}
\usepackage[nameinlink]{cleveref}

\begin{document} 


\baselineskip24pt

\maketitle 

\begin{sciabstract}

%\edB{\gls{CNN} have been established as the main method of image analysis in a wide variety of research fields. We propose its application for the automation of molecular identification through \gls{AFM} imaging with CO-functionalised metal tips. In order to achieve the full generalisation of molecular identification, which arises when dealing the problem with standard classification techniques or even as image captioning, we use a \gls{CGAN} to convert a stack of \gls{AFM} images at various tip-sample distances into a ball-and-stick depiction. CGAN is trained to perform a local mapping of the input image stack onto the ball-and-stick representation, whereby, instead of trying to identify the molecule globally, it correlates the \gls{AFM} image features produced by each atom with a neighbourhood where the structural adsorption configuration plays a key role. To train the network we use QUAM-AFM, the \gls{AFM} image set especially suitable for training neural networks. The models' accuracy has been assessed on both computationally generated and theoretical images, surpassing the previous state of the art regarding the use of \gls{DL} for molecular identification by \gls{AFM}.}

Frequency modulation  (FM) Atomic Force Microscopy (AFM) with metal tips functionalized with a CO molecule at the tip apex has provided access to the internal structure of molecules with totally unprecedented resolution. We propose a model to extract the chemical information from those AFM  images in order to achieve a complete identification of the imaged molecule. Our Conditional Generative Adversarial Network (CGAN) converts a stack of AFM images at various tip-sample distances into a ball--and--stick depiction, where balls of different color and size represent the chemical species and sticks represent the bonds, providing complete information on the structure and chemical composition. The \gls{CGAN} has been trained and tested with the QUAM--AFM data set, that contains simulated AFM images for a collection of 686,000 molecules that include all the chemical species relevant in organic chemistry. Tests with a large set of theoretical images and few experimental examples demonstrate the accuracy and potential of our approach for molecular identification. %(Limited to $~150$ words in Nat. Comm.) \color{red} There are still more than 150.\color{black}

\end{sciabstract}

\newpage

\section{Introduction}

\gls{AFM}~\cite{BinnigPRL1986} in combination with dynamic operation modes~\cite{PerezSurfSciRep2002,GiessiblRevModPhys2003} has become one of the key tools for imaging and manipulation of materials and biological systems at the nanoscale. Operated in the frequency-modulation mode (FM) (commonly known as Non-contact AFM),  \gls{AFM} achieves true atomic-scale resolution~\cite{PerezSurfSciRep2002,GiessiblRevModPhys2003}. The use of metal tips functionalized with a CO molecule at the tip apex, has provided access to the internal structure of molecules with totally unprecedented resolution~\cite{GrossScience2009,PavlicekNatRev2017}. 
The main contrast mechanism for \gls{AFM} with inert tips like CO is Pauli repulsion~\cite{GrossScience2009}, that is due to the overlap of the electron densities of tip and sample. This repulsive force produces positive frequency shifts --changes in the oscillation frequency of the cantilever holding the tip due to the tip-sample interaction-- that are observed as bright features in the constant height \gls{AFM} images above atom positions and bonds, reflecting the molecular structure.
Increasingly accurate \gls{AFM} simulation models~\cite{MollNJP2012, HapalaPRB2014,GuoJPCC2015,SakaiNanoLett2016,EllnerNanoLett2016} have been developed to explain the observed image contrast. They have contributed to elucidate the role of the CO tilting~\cite{HapalaPRB2014}, the influence of other contributions to the tip--sample interaction, like the electrostatic force~\cite{VanDerLitPRL2016,HapalaNatComm2016}, the role of the CO-metal tip charge distribution\cite{EllnerNanoLett2016,EllnerPRB2017}, and the interplay of the short-range chemical interaction and electrostatics in  bond order discrimination and the imaging of intermolecular bonds~\cite{ellner2019molecular}. 

High-resolution experimental (HR) \gls{AFM} images, together with the ability to address individual molecules, have paved the way for the identification of natural products --like breitfussin A, where the structure of some of the fragments was known but methods like NMR failed to provide the global structure~\cite{HanssenACIE2012}--. HR--AFM is also key in the imaging of the intermediates (including radicals) and final products generated in on-surface reactions, shedding light into the formation processes and reaction pathways~\cite{deOteyzaScience2013,KawaiNatComm2016b,kawai2017competing,schulz2017precursor}. 
The technique has been able to resolve more than a hundred different types of molecules in asphaltenes, the solid component of crude oil~\cite{SchulerJACS2015}. 
Molecular identification in all of the previous cases was supported by significant information about the nature of the molecules involved, as in the case of asphaltenes, where we were dealing essentially with polycyclic aromatic hydrocarbons based on C and H atoms. 
In spite of the wealth of information provided by HR-\gls{AFM} experiments and these advances in the interpretation of the observed contrast, the complete identification of molecular systems, i.e. the determination of the structure and composition,  solely based on HR-\gls{AFM} images, without any prior information, remains an open problem.
%%%%%%%%%%%%

% previous works
Few works have tried to tackle this problem using \gls{AI} techniques~\cite{alldritt2020automated,Carracedo2021MDPI} to process AFM images. \gls{DL} is  nowadays routinely used to classify, interpret, describe and analyze images~\cite{krizhevsky2012imagenet,simonyan2014very, Kaiming2016Residual, szegedy2016inception, chollet2017xception, sandler2018mobilenetv2},  providing machines with capabilities that surpass human beings~\cite{he2015delving}. DL ability to recognize patterns could in principle be exploited to characterize the structure of molecular systems.  Gordon \emph{et al.}~\cite{gordon2020automated} implemented a model to automate the detection of spatially correlated patterns in varied sets of AFM images of self--organised nanoparticles. However, the complete atom-by-atom identification posses a significant challenge, as the effects of both geometry and chemical composition contribute to the determination of the 3D molecular charge density, that is ultimately responsible for the AFM contrast.
% Foster
Alldritt \emph{et al.}~\cite{alldritt2020automated} developed a \gls{CNN} whose aim was to determine the molecular geometry from AFM images. The performance was excellent for the structure of quasi-planar molecules, even using the algorithm directly with experimental results. For 3D structures, they were able to recover information for the positions of the atoms closer to the tip, in a height range of 1.5~\AA. However, the discrimination of functional groups produced non conclusive results. 

% Our previous work
In our previous work~\cite{Carracedo2021MDPI}, we showed the feasibility of performing a very accurate automatic molecular classification with \gls{DL} techniques for a set of 60 planar molecules, that include the most common atomic species in organic chemistry, using their  theoretically simulated \gls{AFM} images. Furthermore, we proposed a \gls{VAE}~\cite{kingma2013auto, dilokthanakul2016deep} based method to include the characteristic features of the experimental \gls{AFM} images in the dataset, significantly increasing the accuracy of the model tested with experimental images. However, although this approach shows the potential to recognise both the structure and composition of molecules through \gls{AFM} images, it does not come close to solving the global classification problem, since (i) classification in the usual sense with \gls{CNN} just allows a finite-length output, i.e., only a finite number of structures can be classified, and (ii) we need to consider molecules with a non-planar adsorption configuration.

In this work, we address the problem of molecular identification from a completely new perspective, using visualisation techniques that map images onto images. 
%Our model translates a stack of constant-height AFM images into their ball--and--stick molecular depiction,  the most common graphical representation of molecular systems, that provides a complete identification of structure and composition. \color{red} Repeats a little further down. Delete from  "Our model translates a stack of..." \color{black}
%
Image translation has been widely applied for various purposes, such as image denoising~\cite{zhang2017beyond}, data compression~\cite{rippel2017real, mentzer2018conditional}, synthetic data generation~\cite{oord2016conditional} or image segmentation~\cite{isensee2021nnu}. One of the most widely accepted methods in the community for these tasks is the \gls{CGAN}.  
This enhancement of the original \gls{GAN}~\cite{goodfellow2014generative} has demonstrated an outstanding ability to colorize images, reconstruct objects from edge maps, and synthesize photos from labelled maps, among other tasks \cite{isola2017image}.
%Although this approach has not been used for the specific purpose of identifying molecules,  it has played a key role in problems that could be considered analogous in other fields, such as the fully convolutional translation from aerial photo to map~\cite{isola2017image}. 
In particular, the \gls{CGAN} has played a key role in problems such as the fully convolutional translation from aerial photos to maps~\cite{isola2017image}, that can be considered analogous to our specific goal of molecular identification through ball--and--stick molecular depictions produced from AFM images.

% Architecture of CGAN: suitable for our purposes.
The architecture of a \gls{CGAN} includes two neural networks: the generator and the discriminator. The generator is responsible for converting the input images into the output ones, whereas the discriminator tries to predict %when an image has been produced by the generator and when it is real. 
whether the output image is the real one (ground truth) or has been produced by the generator. The competition between these two networks forces them to improve significantly their performance during the training.
For its prediction, the discriminator compares patches of the generator's input image with its output and with the real image. Thus, these networks specialise in translating and detecting local environments of the images respectively, making the \gls{CGAN} particularly suitable for molecular identification through \gls{AFM} imaging, since the contrast features induced by each atom in the images depend strongly on its chemical environment and very weakly on more distant atoms.

% Our CGAN implementation
In our \gls{CGAN} implementation, the input for the generator is a stack of 10 constant-height HR-AFM images covering the range of tip--sample distances commonly used for AFM imaging, spanning a distance variation of 1~\AA--. % (with the deformations explained in \cref{Sec:IDG_CGAN} used as data augmentation) 
To this end, we have modified the original \gls{CGAN} architecture replacing the 2D convolutions in the first layers of the generator by 3D convolutions that allow processing multiple images. 
Our \gls{CGAN} turns the stack of \gls{AFM} images into a graphical representation, the ball--and--stick depiction,  where balls of different color and size represent the different chemical species and sticks represent the bonds between the atoms,  providing complete information on the structure and chemical composition. The \gls{CGAN} has been trained and tested with the \gls{AFMD}~\cite{QUAM-AFM_repository}, an open-access dataset that includes simulations of theoretical AFM images for a collection of 686,000 molecules that include all the chemical species relevant in organic chemistry. Thus, the model has the ability to identify the structure and composition of any organic molecule, achieving the complete generalisation of the molecular identification problem. 
Below, we discuss the main points of our implementation and test its performance % determined by a visual comparison between prediction and ground truth, 
with a large set of theoretical images and few experimental examples taken from the literature, in order to demonstrate the accuracy and high potential of this approach for molecular identification.

\section{A \gls{CGAN} model to identify molecules through their ball--and--stick depictions} \label{Chap:CGAN}
%\subsection{Conditional Generative Adversarial Network (CGAN)} 

%\subsubsection{Molecular identification with a CGAN}

We use a \gls{CGAN}~\cite{isola2017image} to identify the molecules through ball--and--stick depictions. 
%These images are the most common graphical representation of molecular systems. 
They represent each atomic species with balls of different colours and sizes centered at the position of the atoms, and define the structure through sticks, joining the balls, that represent the chemical bonds. Our proposal is based on the fact that this representation carries chemical information not only in the balls but also through the length of the sticks, since interatomic distances depend on the chemical species and the order of the bond (e.g. single, double and triple C--C bonds have different lengths).

%---------  FIG 1 Our CGAN  -------------------
\begin{figure}[p!]
\centering
\includegraphics[clip=true, width=1.0\columnwidth]{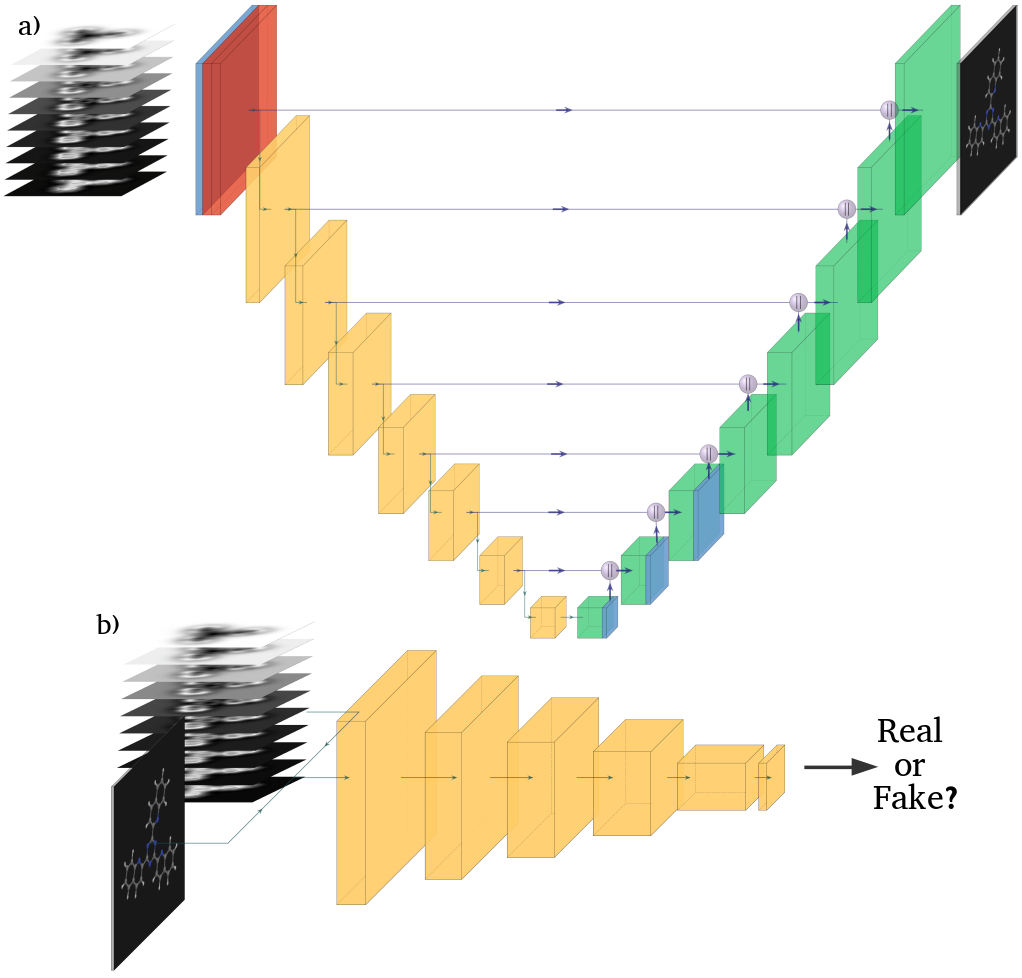}
% \caption{Representation of the \gls{CGAN} structure composed of a generator model (a) and a discriminator model (b). Red boxes represent 3D convolutional layers with batch normalization activated with \gls{lrelu} function, the blue ones represent dropout layers, yellow boxes represent blocks consisting of a 2D convolutional layer with batch normalization activated with \gls{lrelu} and the green boxes represent blocks consisting of a 2D transposed convolutional layer with batch normalization activated with \gls{relu} function except the previous to the output that is activated with a hyperbolic tangent function. A detailed description of each layer can be found in \cref{Sec:A_Model_Details}. }
 \caption{Our implementation of the \gls{CGAN} structure. During the training the generator model (a) and the  discriminator model (b) are confronted against each other in a zero--sum game: firstly, the generator is fed with a stack of \gls{AFM} images and tries to generate the ball--and--stick representation. Secondly, we feed the discriminator with the \gls{AFM} image stack (the same used for the generator) and also with the ball--and--stick depiction. With this data, the discriminator has to predict whether the ball--and--stick depiction is the ground truth or the image generated with the generator network. 
%  Red boxes represent 3D convolutional layers with batch normalization activated with \gls{lrelu} function, the blue ones represent dropout layers, yellow boxes represent blocks consisting of a 2D convolutional layer with batch normalization activated with \gls{lrelu} and the green boxes represent blocks consisting of a 2D transposed convolutional layer with batch normalization activated with \gls{relu} function except the previous to the output that is activated with a hyperbolic tangent function. A detailed description of each layer can be found in \cref{Sec:A_Model_Details}
The models include 3D convolutional layers (red boxes), dropout layers (blue), blocks of 2D convolutional layers (yellow) and  with  2D transposed convolutional layers (green).
For a detailed description of each block and their corresponding layers, including the activation functions, see Methods. %\cref{Sec:CGAN_Details} %\cref{Sec:A_Model_Details}. 
%\edR{For SI: Red boxes represent 3D convolutional layers with batch normalization activated with \gls{lrelu} function, the blue ones represent dropout layers, yellow boxes represent blocks consisting of a 2D convolutional layer with batch normalization activated with \gls{lrelu} and the green boxes represent blocks consisting of a 2D transposed convolutional layer with batch normalization activated with \gls{relu} function except the previous to the output that is activated with a hyperbolic tangent function.}
 }
\label{Fig:CAN}
\end{figure}
%---------  FIG 1 Our CGAN  -------------------

The model applied for the identification is based on the implementation of the \gls{CGAN} proposed in ref.~\cite{isola2017image}. %The model applied for the identification is a modification of the original \gls{CGAN}
%\edR{(RP: Here or INTRO?) This enhancement of the original \gls{GAN}~\cite{goodfellow2014generative} has demonstrated an outstanding ability to colourize images, reconstruct objects from edge maps and synthese photos from labelled maps, among other tasks \cite{isola2017image}.}
%
% Brief description of the CGAN structure
%As shown in \cref{Fig:CAN},
The  \gls{CGAN} model is composed of two networks, known as generator and discriminator.  \Cref{Fig:CAN} shows the structure and layers of each network. We define the stack of 10 \gls{AFM} images at different tip-sample distances as input to the generator and the corresponding ball-and-stick depiction as output. Our proposal differs from the original implementation in the first layers of the generator: a dropout layer with a rate of 0.5 and two 3D convolutional layers (replacing the original 2D convolutional layers) to process the image stack. A dropout layer with  such a high rate is important for the model to be able to generalize and make accurate predictions when dealing with experimental images.

% how a CGAN works...
During the training, the networks are confronted against each other in a zero--sum game consisting of two steps. Firstly, the generator is fed with a stack of \gls{AFM} images and tries to generate the ball--and--stick representation corresponding to the molecule from which the input \gls{AFM} images have been simulated. Secondly, we feed the discriminator with the \gls{AFM} image stack (the same used for the generator) and also with the ball--and--stick depiction. With this data, the discriminator predicts whether the ball--and--stick depiction is the ground truth or %the image generated with the generator network by analysing the inputs segmented into patches. 
the image generated with the generator network. %  by comparing the inputs segmented into patches of 16$\times$16 pixels. 
%Notice that, in order to prevent the discriminator from learning that the ball--and--stick representation is always generated by the first network, it is alternately fed with the ground truth and with the output of the generator. 
%It should be noted that the discriminator does not directly compare the input of the generator with the output, but rather patches of the image which the original paper defines as PatchGAN. This makes \gls{CGAN} especially powerful in \gls{AFM} image analysis, as the resulting effects on the images depend strongly on the local environment and smoothly on the global structure.
%
%% Technical details for training
%
%In order to force the generator, not only to fool the discriminator, but also to produce outputs  close to the real ones and with as little blur as possible, the distance L1 is added to the loss function weighted with the factor $\lambda=100$ defined by Isola~\textit{et al.}~\cite{isola2017image}. 
%
%
%In this way, we train the two networks together in a end-to-end process in which the first network learns to generate images as close as possible to the ball--and--stick depiction, 
%In this way, we train the two networks together in a end-to-end process in which the first network learns both to fool the discriminator and to generate images as close as possible to the ball--and--stick depiction in a L1 sense and the discriminator learns to guess whether the second input image is real or fake. 
% Training
In this way, we train the two networks together in a end-to-end process in which the first network learns both to fool the discriminator and to generate images as close as possible to the ball--and--stick depiction, and the discriminator learns to guess whether the second input image is real or fake. 
From a practical point of view, the discriminator is a network that is only useful to force the generator to improve. Therefore, once this objective has been achieved, we discard the discriminator network. The generator is in charge of generating the ball--and--stick depiction representing the atoms and bonds, providing a complete identification of the molecule. 

% two important technical points 
While most of the model details are presented in the Methods section, there are two  technical points that we want to highlight as they are important in order to explain the remarkable performance of our method approach.
The first one is related to how the discriminator makes its prediction. This is not achieved by a global assessment of the inputs but by comparing them segmented into patches of 16$\times$16 pixels. This local analysis based on small patches of the images makes \gls{CGAN} especially powerful in \gls{AFM} image analysis, as the features induced by the structure and composition  on the AFM images depend strongly on the local chemical environment and smoothly on the global molecular configuration.
The second one exploits the freedom to incorporate additional terms into the loss function used during the training. As suggested in the original \gls{CGAN} implementation~\cite{isola2017image}, a distance L1 (defined as the sum of the absolute difference of the components of a vector) has been added to the loss function. This distance, an alternative to the usual Euclidean L2 norm, forces the generator not only to fool the discriminator, but also to produce outputs closer to the real ones and with as little blur as possible. %\color{red} The loss function is part of the CGAN, I do not understand what this paragraph means. \color{black}

%\edR{Move to SI: The generator of the \gls{CGAN} was compiled with \gls{MAE} (using the parameter $\lambda=100$ defined by Isola~\textit{et al.}~\cite{isola2017image}) ,  while the binary cross entropy was used for the discriminator. The model was minimised by applying batches of 32 inputs with \gls{Adam} optimiser where the learning rate and first moment parameters were set to $2\cdot 10^{-4}$ and $0.5$ respectively. The training of the model was carried out during six epochs (109K iterations), displaying 300 predictions of the validation set to estimate the optimal training point every 10.000 iterations.}

\section{Results} \label{Sec:results}

\subsection{Testing the identification with simulated AFM images} \label{Sec:Simulated_Test}

%---------  FIG 2 perfect predictions -------------------

\begin{figure}[b!]
\centering
\includegraphics[clip=true, width=1.0\columnwidth]{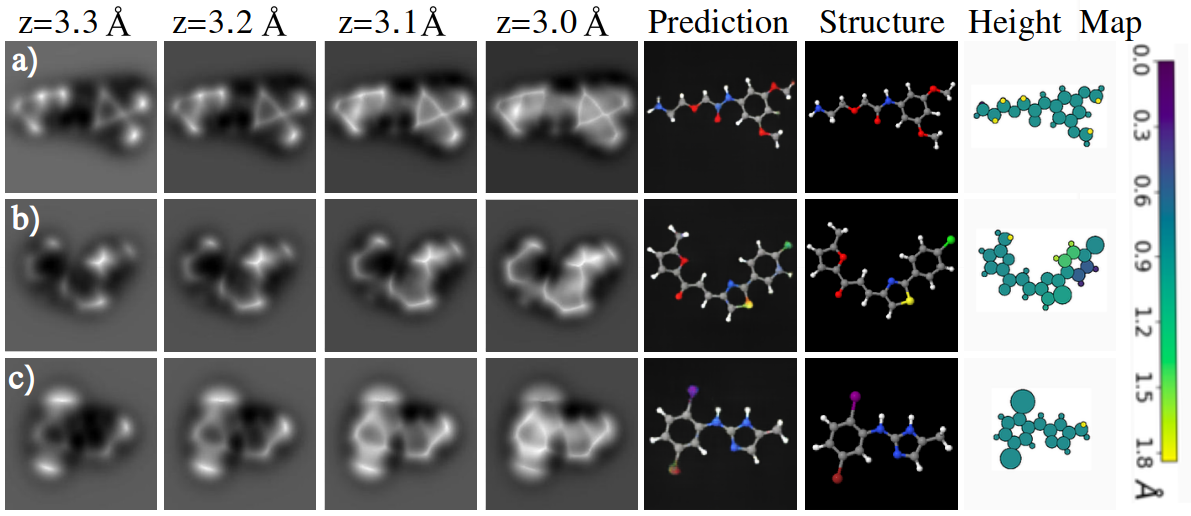}
 \caption{(From left to right) \gls{AFM} images at different tip-sample distances, prediction performed by the \gls{CGAN}, real structure and height map for (a) 2-(2-aminoethoxy)-N-(3,5-dimethoxyphenyl)acetamide, (b) 3-[2-(4-chlorophenyl)-1,3-thiazol-4-yl]-1-(5-methylfuran-2-yl)prop-2-en-1-one and (c) N-(5-bromo-2-iodophenyl)-5-methyl-1H-imidazol-2-amine.}
\label{Fig:Th_Test1}
\end{figure}
%---------  FIG 2 perfect predictions  -------------------

In order to evaluate the accuracy of molecular identification through \gls{AFM} with the \gls{CGAN}, we perform a test with 3.015 structures randomly selected from the set of 81K molecules specifically reserved for this purpose from \gls{AFMD} (see Methods). The test was not performed on the complete test set due to the fact that the evaluation was carried out by human visual comparison between the target structure and the one predicted by the model. For each of these structures, we randomised the selection of the simulation parameters among the 24 possible combinations offered by \gls{AFMD} (see Methods), resulting in 3.015 stacks of 10 tip-sample distance \gls{AFM} images. % with the range of tip--sample distances defined in \cref{Sec:Simulation_Parameters}. 

%---------  FIG 3 accuracy vs molecular corrugation -------------------

%\begin{table}[b!]
\begin{figure}[b!]
\centering
% \begin{tabular}{| l| c| c | c | c |}
% \hline
% \hline
% z diff.& Support& Acc.Tot.& Struct.Acc.& Acc. Atoms  \\ 
% \hline
% \hline
% [0,183] pm & 3015 & 0.74 & 0.95 & 0.96 \\  
% \hline
% [0,50) pm & 2301 & 0.84 & 0.98 & 0.98 \\  
% \hline
% [50,100) pm & 294 & 0.56 & 0.90 & 0.92 \\  
% \hline
% [100,150) pm & 378 & 0.32 & 0.82 & 0.86 \\  
% \hline
% [150,183] pm & 42 & 0.20 & 0.60 & 0.82 \\  
% \hline   
% \end{tabular}
%\includegraphics[clip=true, width=1.0\columnwidth]{./IMAGES/Histograms/All_histograms.png}
%\includegraphics[clip=true, width=1.0\columnwidth]{../../IMAGES/Histograms/bars_diagram.png}
\includegraphics[clip=true, width=1.0\columnwidth]{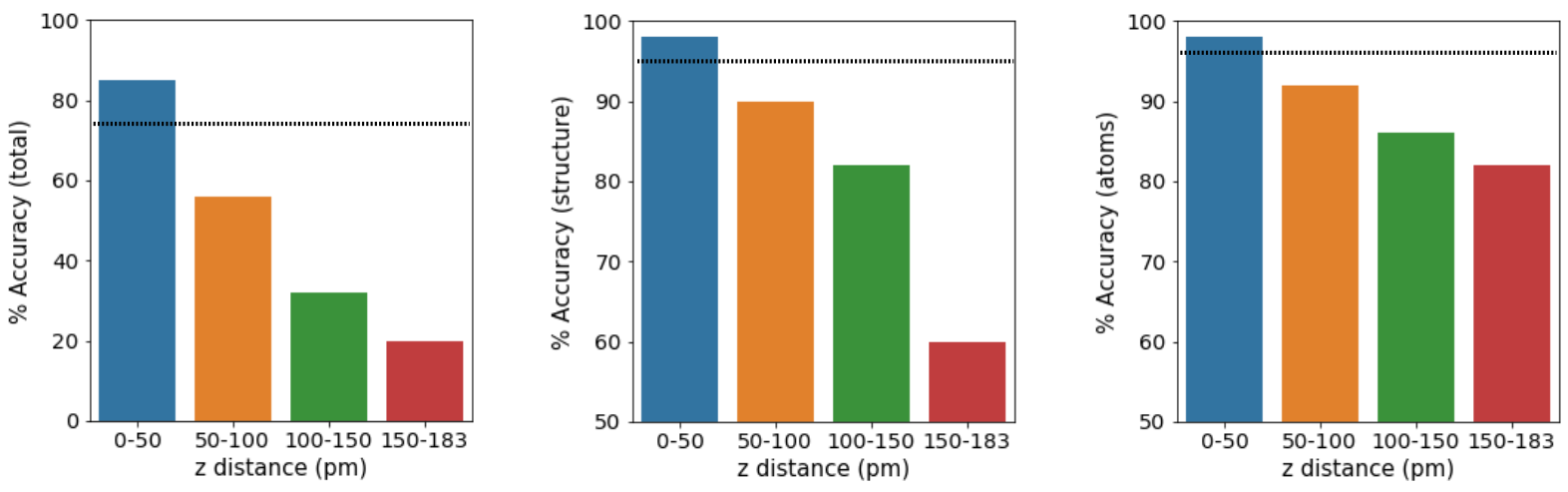}
\caption{Accuracy of the model in a test where both the 3015 structures and their simulation parameters have been randomly selected.  The bar charts show (from left to right) the overall accuracy (perfect structure and atom prediction), the accuracy of structure discovery, and the accuracy in revealing the atomic species. The set of structures has been divided into four subsets according to their torsion in order to show the dependence of the model accuracy versus the height difference in the atoms of the molecule. The horizontal dashed line shows the accuracy over the complete test set. The (total) accuracy has been evaluated considering that the final result is correct only if  the prediction is perfect: it shows all the bonds of the molecule, the number of vertices of each structure (chain or rings), and the proper color assigned to each atom, with the exception of the hydrogens and its bonds.  %it shows all of the bonds of the molecule and the number of vertices in each structure (chain or rings), with the exception of the hydrogens and its bonds. 
The structure accuracy has been calculated as the percentage of fully discovered (perfect) structures out of the total set of structures. The accuracy in the prediction of the atomic species has been evaluated as the percentage of total hits (correct predictions) over the total number of atoms in the set, without considering the hydrogens. See table S1 for details.}
%\caption{Accuracy of the model in a test where both the 3015 structures and their simulation parameters have been randomly selected.  The table shows (from left to right) the range of maximum height differences between the atoms of each molecular structure in the evaluated ensemble, the number of structures in the evaluated set, the overall accuracy (perfect structure and atom prediction), the accuracy of structure discovery, and the accuracy in revealing the atomic species. The accuracy of the structures has been evaluated considering that the final result is correct only if  the prediction is perfect: it shows all  of the bonds of the molecule and the number of vertices in each structure (chain or rings), with the exception of the hydrogens and its bonds. The structure accuracy has been calculated as the percentage of fully discovered (perfect) structures out of the total set of structures. The accuracy in the prediction of the atomic species has been evaluated as the percentage of total hits (correct predictions) over the total number of atoms in the set, without considering the hydrogens. \edR{mover la tabla a la SI y añadir una linea horizontal en cada histograma con la accuracy para el conjunto completo}}
\label{Table:Accuracy_Th}
\end{figure}
%---------  FIG 3 accuracy vs molecular corrugation -------------------

%The main point revealed by \cref{Table:Accuracy_Th} is that the \gls{AFM} images do indeed contain sufficient information to carry out both chemical and structural identification of the molecule. Although this prediction has been performed on a theoretically simulated test set, it should be noted that there are structures that a human expert would not be able to reveal (see \cref{Fig:Th_Test1}), so the proposal to apply a \gls{CGAN} to molecular identification through \gls{AFM} images is an approach with outstanding results. Overall, the model recognises both chemically and structurally semi--flat molecules in complex cases. \Cref{Fig:Th_Test1}~(a) shows the identification of 2-(2-aminoethoxy)-N-(3,5-dimethoxyphenyl)acetamide, whose \gls{AFM} images are characterised by strong contour distortions created by the charge distribution of the oxygens, which, on the one hand, hide the bonds of the \textit{sp3} carbons and, on the other hand, mitigate the response of the nitrogens in the centre of the chain. In addition, the ability to differentiate other substituents such as \textit{sp3} and \textit{sp2} carbons and amino groups should be highlighted. \Cref{Fig:Th_Test1}~(b and c) shows other notable identifications of the model, such as the identification of three different halogen species (Ch in \cref{Fig:Th_Test1}~(b) and I and Br in \cref{Fig:Th_Test1}~(c)), \textit{sp3} carbons, sulphur, and oxygens and nitrogens with different surroundings.

% 1. we show that the model works, beyond human capabilities.
The results of the test shown in \cref{Fig:Th_Test1} demonstrate that our method works with outstanding results: theoretically simulated \gls{AFM} images contain sufficient information to carry out a complete chemical and structural identification of the molecule through the prediction of its ball--and-stick prediction.
The model recognises both chemically and structurally semi--flat molecules in complex cases, including structures that a human expert would not be able to identify.
\Cref{Fig:Th_Test1}~(a) shows the identification of 2-(2-aminoethoxy)-N-(3,5-dimethoxyphenyl)acetamide, one of these tough examples. 
The corresponding \gls{AFM} images are characterised by strong distortions of the structure created by the strong charge accumulation around the oxygens~\cite{zahl2021TMA}. These strongly electronegative atoms hide their bonds with the \textit{sp3} carbons, creating a triangular feature at the position of the ring and hiding also the presence of the N atom attach to it. Nevertheless, the model is able to differentiate  \textit{sp3} and \textit{sp2} carbons and identify the two amino groups, leading to a perfect prediction. \Cref{Fig:Th_Test1}~(b and c) shows other remarkable achievements of the model, such as the identification of \textit{sp3} carbons, sulphur, oxygen and nitrogen atoms in different chemical environments and the accurate discrrimination of three different halogen species (Cl in \cref{Fig:Th_Test1}~(b) and I and Br in \cref{Fig:Th_Test1}~(c)).

\Cref{Table:Accuracy_Th} provides a quantitative estimate of the accuracy of our identification method using a global assessment  and two specific evaluations focused on either structure or composition.  The model achieves a remarkable 74\% of perfect predictions, that increase to 95\% (96\%) when considering only structure (composition). Notice that, in the total accuracy and the structure accuracy, a prediction has been considered correct only if there is a perfect match,  whereas the accuracy in the prediction of each atomic species has been assessed by considering each individual atom in the molecule as correct or incorrect. This method of evaluation penalizes errors in structure discovery more than in atom determination, since in all the predictions most of the structure is revealed correctly, providing valuable information about the molecule, in spite of been considered as incorrect in the determination of the accuracy.  

% 3. Influence of the molecular corrugation in the accuracy.
We have explored the influence of the molecular corrugation --the maximum height difference of the atoms in the molecule (excluding hydrogens), where the height is defined as the distance between atoms measured perpendicular to the molecular plane--., in the performance of the model.  
The force curves associated with certain atomic species in different molecular moieties are quite similar.   In fact, in some cases, these curves are almost identical except for a rigid translation, equivalent to a vertical displacement of the atoms. Thus, we could expect the model to mistake some of these atoms in non--planar structure where they are at different heights.
The test set was split into four subsets according to the maximum height difference and the accuracy was evaluated independently for each subset. According to \cref{Table:Accuracy_Th}, both the total  and the composition accuracy 
decrease linearly with the maximum height difference, while the structure accuracy shows this linearly behavior in the range [0,1.5] \r{A} but has a stronger decay from 1.5 \r{A} onwards. 

%\begin{figure}[b!]
%\centering
%\includegraphics[clip=true, width=1.0\columnwidth]{../../IMAGES/TH_TEST/CUALITATIVO/TH_PRED3.png}
% \caption{\gls{AFM} images, predictions and structures of (a and b) 2-quinolin-8-ylisoindole-1,3-dione, (c and d) 4-N-(2-bromophenyl)-2-N-phenylpyridine-2,4-dicarboxamide, (e and f) 3-amino-2,6-difluoro-N-(2,3,4-trifluorophenyl)benzamide. (a, c and e) show the results of molecules in gas phase while (b, d and f) are their respective results of relaxing the structure but imposing that the atoms remain in the same plane.}
%\label{Fig:Th_Test3}
%\end{figure}
%---------  FIG 4 understanding the errors -------------------
\begin{figure}[b!]
\centering
\includegraphics[clip=true, width=1.0\columnwidth]{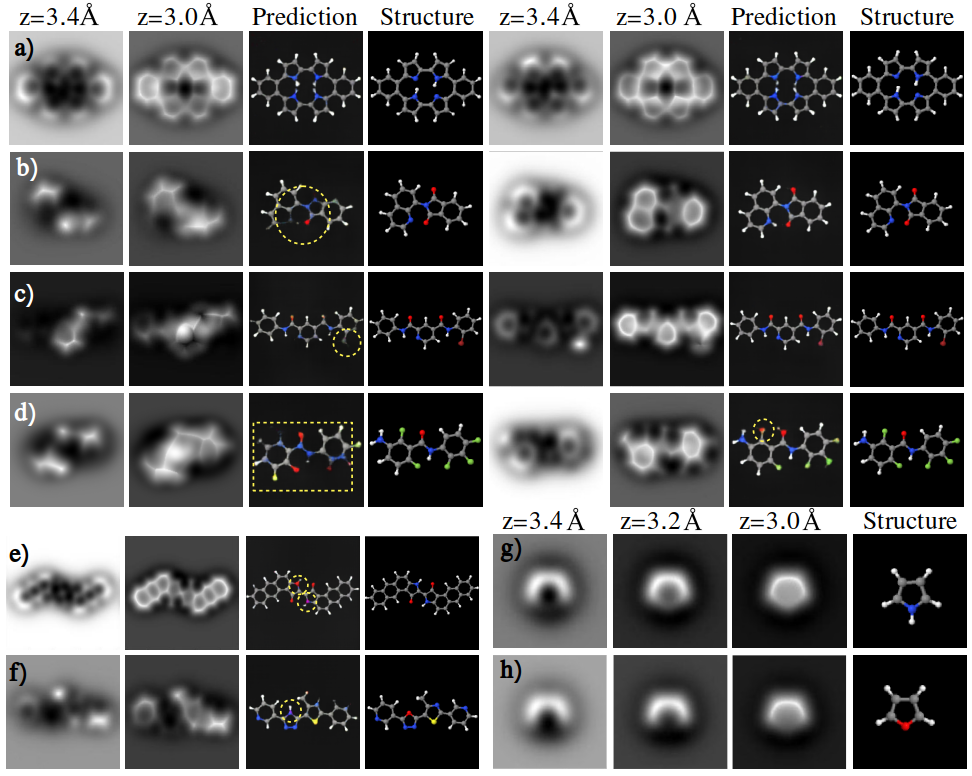}
%\caption{\gls{AFM} images, predictions and structures of (a) \gls{mDBPc} relaxed on Ag(111) surface and a NaCl bilayer. Predictions of the gas-phase molecules (left) and with all atoms in the same plane (right) of (b) 2-quinolin-8-ylisoindole-1,3-dione, (c) 4-N-(2-bromophenyl)-2-N-phenylpyridine-2,4-dicarboxamide and (d) 3-amino-2,6-difluoro-N-(2,3,4-trifluorophenyl)benzamide. Predictions of (e) 2-(1-oxo-3H-naphtho[2,3-e]indol-2-ylidene)-3H-naphtho[2,3-e]indol-1-one and (f) 2-(4-methyl-2-pyridin-3-yl-1,3-thiazol-5-yl)-5-pyridin-3-yl-1,3,4-oxadiazole in gas-phase. (g and h) Comparison of \gls{AFM} images of pyrrole and furan.}
 \caption{(a) \gls{AFM} images, predictions and structures for \gls{mDBPc} relaxed on a Ag(111) surface (left) and on a NaCl bilayer (right), see ref.~\cite{shimizu2020effect} for structural torsion details on each surface. \gls{AFM} images, predictions and structures in the gas-phase configurations (left) and in a forced planar structure (right) for the molecules (b) 2-quinolin-8-ylisoindole-1,3-dione, (c) 4-N-(2-bromophenyl)-2-N-phenylpyridine-2,4-dicarboxamide and (d) 3-amino-2,6-difluoro-N-(2,3,4-trifluorophenyl)benzamide. \gls{AFM} images, predictions and structures for (e) 2-(1-oxo-3H-naphtho[2,3-e]indol-2-ylidene)-3H-naphtho[2,3-e]indol-1-one and (f) 2-(4-methyl-2-pyridin-3-yl-1,3-thiazol-5-yl)-5-pyridin-3-yl-1,3,4-oxadiazole in a gas--phase configuration. (g and h) Comparison of \gls{AFM} images of pyrrole and furan.}
\label{Fig:Th_Test2}
\end{figure}
%---------  FIG 4 understanding the errors -------------------
 
% 4. We try  to understand the  most common mistakes and the accuracy vs molecular corrugation
\Cref{Fig:Th_Test2} provides some important hints on the origin of the limitations of the model revealed by the statistical analysis presented above. 
Starting with the role of the maximum height difference, the left panels in \cref{Fig:Th_Test2}~(b, c and d) show two representative \gls{AFM} images, the prediction and the real structure for three molecules that have a strong torsion in their gas--phase configuration. These images show that the model perfectly identifies chemically and structurally the top part of the molecules,  but fails with the bottom, where the CO tip cannot retrieve enough information during constant height imaging, even at the shortest tip-sample distances, due to the CO lateral relaxation. These results explain the lower accuracy of the model for the molecules with stronger torsion, particularly in the case of the structure accuracy, that requires a perfect identification of the whole molecular structure. At the same time, it seems to confirm that there is a limit beyond which it is not possible to obtain information from an \gls{AFM} with the current operation setups and with a single adsorption orientation of the molecule~\cite{alldritt2020automated}.

We do not expect this limitation to be so crucial when dealing with a molecular identification based on experimental images, where the molecules are deposited on a substrate. The final  adsorption configurations are significantly flatter than the gas--phase ones, as the attractive molecule-substrate interaction compensates the steric hidrance effects responsible for the  torsion, even in the low reactive substrates commonly used for \gls{AFM} experiments. This idea has been tested with the three molecules in \cref{Fig:Th_Test2}~(b, c and d). The left panels of \cref{Fig:Th_Test2}~(c) show that in the gas phase structure,  the model correctly predicts that bromine is a halogen (by bond length and ball size) but does not determine the color of the ball. A similar case is presented in \cref{Fig:Th_Test2}~(d), where several atoms are misclassified. We have forced these three molecules to acquire a flat structure. The corresponding AFM images, the new prediction and the structure are shown on the right panels of \cref{Fig:Th_Test2}~(b, c and d). %\color{blue} In the right panels of \cref{Fig:Th_Test2}~(b, c and d), we show the \gls{AFM} images and the corresponding prediction of these molecules simulated from a planar configuration. \color{black} 
The prediction becomes perfect with respect to the structure in all of the three cases, and, composition-wise, fails only in a single atom in the case displayed in \cref{Fig:Th_Test2}~d. % (\edR{there is a mistake also in \cref{Fig:Th_Test2}~b where a N in the lower hexagonal ring is predicted as N--H. Hacer un comentario en algun sitio sobre el problema especifico con los H, en particular cuando van asociados a los N}) .  

% 5. so, why training with torsion... ? 
After the analysis presented above, it is sensible to ask if the choice of training the model with the structures in \gls{AFMD}, that correspond to  gas--phase configurations, is the best option for molecular identification based on experimental images. This choice have been taken in the first place to make the simulation computationally feasible, as it is simply not possible to perform the relaxations needed to determine the adsorption configurations of all the molecules in the data set on a number of different substrates. 
However, our choice, more that a practical consideration, is actually guided by the fact that the AFM contrast of the different chemical species is strongly influence by the chemical environment. 
Training the model with the molecular structures in \gls{AFMD}, that, in general,  do not correspond to the adsorbed configuration in the experiments, is 
providing the model with the necessary information to learn the local relationships that the different chemical species may have depending on the height. Instead of learning to identify a structure in one particular configuration, the model is learning to relate atoms to their surroundings, allowing it to recognise molecules in different configurations.

% the example: mDBPc
\Cref{Fig:Th_Test2}(a) demonstrates  this idea. It shows the \gls{AFM} images calculated for the stable adsorption configuration of  \gls{mDBPc} on two different substrates: a more reactive Ag(111) surface and a rather inert NaCl bilayer. The final structures are quite different and neither of them is flat. This reflect in the different AFM contrast, that is in excellent agreement with the experiments~\cite{shimizu2020effect}.
When the stack of images corresponding to these two configurations is shown to our model, the prediction for the structure and composition of the molecule is perfect in both cases, except for the position of the two internal hydrogens that are always very difficult to determine from AFM experiments.
%\edR{except for the position of the two internal hydrogens, that can actually be interchanged among the four N atoms on the sides of the inner ring, leading to tautomeric configurations that cannot be precisely determined from the  AFM experiments. (Or simply say that H are very difficult to identify in AFM???)}
%  
This example with theoretical images and the experimental cases discussed below show that the training with the highly corrugated gas--phase configurations, although not enough to keep its global accuracy in the tests performed with molecules with strong torsions, is actually an important asset of the model. 
These structures are making the model robust by showing how features associated with atomic species and molecular moieties evolve with the variation of height in different chemical environments. 
The choice of the molecular adsorption configurations on a particular substrate for training may lead the model to specialize excessively and loose the ability to generalize and identify the same molecule adsorbed on a different substrate. The gas--phase structures, combined with the choice of images generated with different \gls{AFM} operational parameters and the use of an \gls{IDG} (see Methods), introduce enough variability during the training to allow the model to identify the molecule, despite the differences introduced by the substrate. 
In summary, the torsion of the gas--phase structures, rather than being a limitation, is enhancing the ability of the model to generalize and to recognise molecules in different adsorption configurations.

Beyond the subtleties in the AFM contrast created by the interplay of the chemical nature of the atoms,  their chemical environment and their relative height, we have identified some misclassifications that occur with some frequency, even in rather flat configurations. \Cref{Fig:Th_Test2}~(e and f) shows two examples where  the model swaps a N--H group in a pentagon for an O atom. %(\edR{RPP: we have to mark also the other N--H that is mistaken by something like O--H}) %It is assumed that H is practically unidentifiable by \gls{AFM}. 
In this case, although chemically they have different properties, the fact that the atoms are very electronegative and have a similar charge distribution reflects in the similar features they show in the \gls{AFM} simulations in a perfectly planar configuration (see \cref{Fig:Th_Test2}~(g and h)). This fact makes them extremely difficult to identify in the presence of small variations in height. 
Another pair that is frequently mistaken for variations in height is  
%\rpp{the O with F connected to an aromatic ring} 
O and F when connected to an aromatic ring (see \cref{Fig:Th_Test2}~(d)). This case is more surprising since, even though the two atoms are highly electronegative and of similar size, the O double bonded to a C of an aromatic ring should, at first, show some distinctive feature with respect to a C--F. Although the F and O features are similar, one would expect them to be distinguishable in a planar structure. It is not clear whether this error is due to some unknown effect on the structure or, perhaps, as they have similar sizes in the ball--and--stick representation, the model mistakes them under certain conditions.

\subsection{Molecular identification based on experimental AFM images}

The final goal of our \gls{CGAN} model is to identify molecules from their experimental \gls{AFM} images.
As discussed above, the range of \gls{AFM} operational parameters used to simulate the images generated for each of the molecules and the use of gas--phase configurations  introduce enough variability during the training to allow the model to identify the molecule, despite the differences introduced by the substrate. We have explicitly tested this point with theoretical AFM images generated for the adsorption configurations of mDBPc on two different substrates with quite different reactivity, a Ag(111) surface and a NaCl bilayer  (see \cref{Fig:Th_Test2} (a)).  The theoretical AFM images faithfully reproduced the experimental results~\cite{shimizu2020effect}.

\begin{figure}[pt]
\centering
\includegraphics[clip=true, width=1.0\columnwidth]{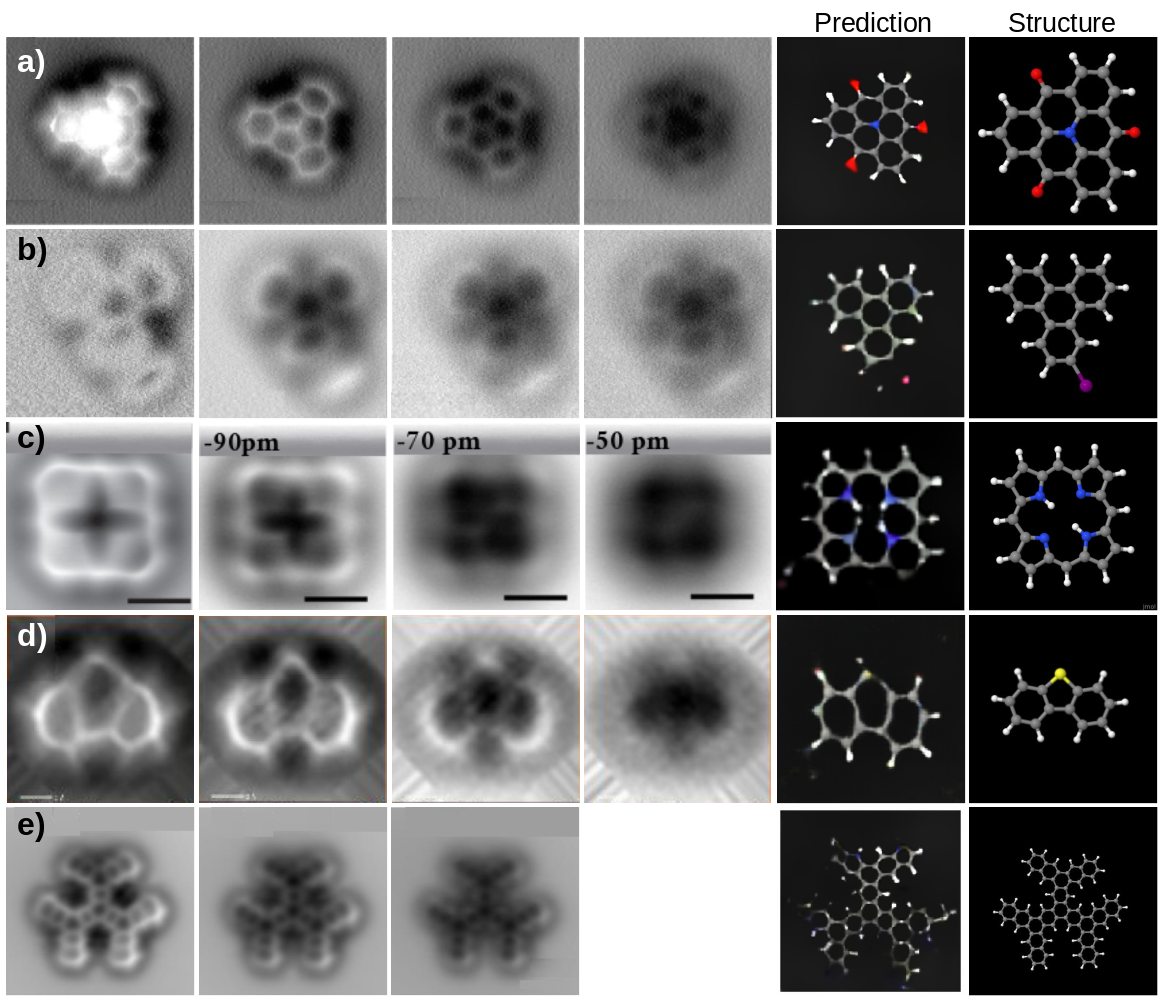}
\caption{Experimental \gls{AFM} images, prediction performed with the \gls{CGAN} and structure for (a) 1-azahexacyclo[11.7.1.1\textsuperscript{3,19}.0\textsuperscript{2,7}.0\textsuperscript{9,21}.0\textsuperscript{15,20}]docosa-2,4,6,9(21),10,12,15,17,19-nonaene-8,14,22-trione, (b) 2-iodotriphenylene, (c) 21,23-dihydroporphyrin, (d) dibenzothiophene and (e) [19]dendriphene.  Experimental images in (a, c, d, e)~\cite{HeijdenACSNano2016, fatayer2019molecular, EXP, vilas201819} were taken in the FM operation mode at constant height, while, in (b)~\cite{martin2020bond}, a novel Q-control \gls{AM-AFM} mode was used.
The color code for the balls representing the chemical species is: carbon (grey), hydrogen (white), oxygen (red), iodine (purple) and nitrogen (blue). From top to bottom, 4, 6, 6, 10 and 3 images were published, so we have linearly interpolated these images in order to produce the 10--image stack used by our \gls{CGAN} model as input (see text and Figs.~S1 and S2 for details). Images are reproduced by courtesy of the American Chemical Society (ACS), American Physical Society (APS), American Association for the Advancement of Science (AAAS), and European Chemical Societies Publishing.}  
\label{Fig:Exp_Test}
\end{figure}

% 2.Our disclaimer
Now, we want to assess the accuracy of the model with experimental results. This test is going to be limited by the scarce number of published AFM studies that include sets of images as a function of the tip height. Furthermore, most of these few studies neither provide sufficient images  (10 images, taken at 0.1~\AA\ intervals) nor are in the range of tip--sample distances (2.80--3.70 \AA) which our analysis with simulated images have shown necessary to properly sample the variation of the tip--sample interaction and achieve complete chemical identification. Despite these drawbacks, the results presented below are really promising.  
To test the performance of the model with experimental results,  we have selected sets of \gls{AFM} images originally published in refs~\cite{HeijdenACSNano2016, martin2020bond, fatayer2019molecular, vilas201819, EXP, martin2019bond}. In general, fewer than ten images corresponding to different  tip--sample distances were published in these papers, so we have linearly interpolated the images two by two to extract additional images to complete the input, the stack  of 10 images, required for the \gls{CGAN} model. In some cases the experimental results were so limited, that it was necessary to weigh differently each image to obtain multiple results from each image pair (see \cref{Fig:Exp_Test}  and figs.~S1 and S2). %Pred_Experimental_Iodothiph}). 
We have denoised the generated 10-image stack by applying the \textit{medianBlur} filter with size 3 from the OpenCV Python package.

It is important to stress that the interpolated images are generated for the sole purpose of completing the input dimensions required by the model, i.e. they do not provide additional information to that supplied by the original images. Therefore, the test with experimental images is really tough: We are not only increasing the complexity by using as inputs experimental images --simply cut and edited from different publications and that, in spite of the applied filter, always carried some noise--, but we are also severely reducing the amount of information with which we feed the model. %we test the model by reducing the input data by half and also by reducing the quality thereof as they are experimental images, which, although of high resolution, show relevant differences with the theoretical simulations \color{red} ref a classification results\color{black}. 

%---------  FIG 5 experimental test -------------------
%---------  FIG 5 experimental test -------------------

% robust against tip asymmetries 
A drawback that may hinder chemical identification by experimental \gls{AFM} imaging is that the observed interaction depends on the details of the tip structure, like the attachment of the CO molecule to the metal tip.  \Cref{Fig:Exp_Test}~(a) shows experimental \gls{AFM} images, taken at constant height and acquired with a CO--terminated tip, for a 1-azahexacyclo[11.7.1.1\textsuperscript{3,19}.0\textsuperscript{2,7}.0\textsuperscript{9,21} .0\textsuperscript{15,20}]docosa-2,4,6,9(21),10,12,15,17,19-nonaene-8,14,22-trione molecule adsorbed on a Cu(111) surface. These \gls{AFM} images (and, by inheritance, also their interpolations) show an imperfect threefold symmetry.   Although this asymmetry could be related to the adsorption configuration of the molecule, the discussion in ref.~\cite{HeijdenACSNano2016} proves that it is really caused by the flexibility of the CO--Cu bond coupled with an asymmetric tip.  Therefore, the chemical identification of this molecule has two additional complications,  besides the lack of input data and the switch to experimental images: First, this structure is not part of the training set, so, in addition to testing the model with an experimental image, this is a perfect example to verify its ability to generalise. On the other hand, because in the theoretical simulations tip irregularities are not considered, % beyond the lateral tilting stiffness,
 the model has not been trained with images containing characteristic features induced by these asymmetrical tips in the experimental images. Despite these drawbacks, the \gls{CGAN} is not only able to reveal the molecular structure but also to predict with perfect accuracy the chemical species that make up the molecule. %Therefore, in addition to the simple identification of the structure, our model is able to predict with high accuracy the chemical species that make up the molecule. 
 
 % better performance with chemical composition
Besides been robust against tip asymmetries, the model seems to perform better in the determination of the chemical composition with experimental images. As discussed above (see \cref{Sec:Simulated_Test}), one of the most common errors in the tests performed with simulated images was to mistake O for F in complex molecules, as they produced a similar AFM contrast. However, in the prediction of this molecule through the experimental \gls{AFM} images, where the symmetry is affected by the irregularity of the tip, the model identifies the three oxygens with absolute accuracy (see \cref{Fig:Exp_Test}~(a)). It is not possible to make a general statement since the test with oxygens is limited to their presence in this particular structure, but this result seems to indicate that our \gls{CGAN} is able to clearly differentiate % better
some chemical species, like oxygens and fluorines, in experimental images. %\color{blue} In general the cgan will identify better chemical species in simulated images, which is what we have trained it with. The same applies to the first sentence of this paragraph. \color{black}

Our \gls{CGAN} model seems to work also with constant-height images taken using different AFM operation modes.
\Cref{Fig:Exp_Test}~(b) shows the prediction performed for  2-iodotriphenylene on Ag(111) with a stack of \gls{AFM} images taken using the measured oscillation amplitude in a new operation mode, Q--controlled \gls{AM-AFM} with CO--functionalized tips operated in constant--height mode, proposed in ref.~\cite{martin2020bond}. %(\href{https://doi.org/10.1063/5.0018246}{Please, check Figure S4 of supplementary material of this paper.}) 
The \gls{AFM} images resulting from using both phase modulation in Q-control \gls{AM-AFM} and frequency modulation (FM) modes on the same molecule as well as the respective predictions performed by the model are shown in fig.~S1. %(\edR{we should include also the inputs and interpolated images for Q--control mode in this figure.}) \color{blue} Q-control images are shown in \cref{Fig:Exp_Test}. Is the proposal to duplicate the image by adding the interpolated images in the SI? If we do it for that image maybe we should do it for all the ones in \cref{Fig:Exp_Test}. In my opinion the interpolated images do not add much information. \color{black}%\cref{Fig:Pred_Experimental_Iodothiph}. 
% Fig S1
As discussed in section S2, none of these AFM images correspond to the AFM operation mode used to simulate the AFM images employed in the training of the model. This is clear for the amplitude  (\cref{Fig:Exp_Test}~(b)) and phase images, but it is also the case in the FM images, as the oscillation amplitude is very different (varying from 45 to 525 pm) in each of the experimental  images, while the10--image stacks used in the training correspond to different tip--sample distances of closest approach but to the same oscillation amplitude. Moreover,  the tip--height range covered by the images (64 pm) is significantly smaller than the 100pm that we consider optimal and has been chosen so that similar contrast features were shown in the amplitude, phase and FM images. Finally, we have included in our analysis the amplitude image at the closest distance, that shows a significantly different contrast.

In spite of these severe limitations in the input, the model fed with the amplitude images fully reveals the molecular structure and the presence of the I atom.  
In the case of phase and FM images, the model gives a good description of the molecular structure but fails to provide a clear prediction about the halogen, since the color is more like the one associated to bromine than the one  corresponding to iodine (See fig.~S1).   
Far from considering these predictions a failure, these results indicate that our CGAN model can provide very useful information regarding the molecular  identification when fed with images taken with different AFM operation modes.
% Fig S2
Nevertheless, more work is needed to reach a final conclusion about the merits and limitations of our model for this particular case, 2-iodotriphenylene on Ag(111), as shown by the analysis of another series of constant--height images taken in the frequency modulation mode for the molecule and for the products of a dehalogenation reaction locally triggered using a voltage:  a triphenylene (TP) radical and the cleaved I atom~\cite{martin2019bond} (see section~S2).  The image features at the halogen position and its evolution with tip height in Fig.~S2(a) are quite different from those shown in other experimental examples and from our AFM simulations, and the model predicts a methyl group instead of a halogen.  In the case of the dehalogenation products (Fig.~S2(b)), our model captures the presence of the cleaved I atom and provides a strongly deformed structure where the dehalogenated ring is not closed, consistent with the lack of information in the AFM images due to the strong bending of the molecule towards the substrate induced by the interaction of the unsaturated C bond in that ring with the metal.

% another successful tests
\Cref{Fig:Exp_Test}~(c) shows another rather successful identification, in this case, a 21,23--dihydroporphyrin molecule. The test has been carried out with interpolations from five experimental images that cover tip--sample distances varying in a range of 1~\AA, although the average distance seems to be larger than the one used in the simulations of \gls{AFMD}. The model is able to reveal the four pentagonal rings and the position of the nitrogens.

%\begin{figure}[pt]
%\centering
%\includegraphics[clip=true, width=0.8\columnwidth]{../../IMAGES/EXPERIMENTAL_TEST/Predictions4.png}
% \caption{(a) A series of force image slices taken each 10pm (from left to right and from top to bottom) of a dibenzothiophene molecule, the ball--and--stick depiction (first row) and the prediction performed by the \gls{CGAN} (second row). The \gls{AFM} images were originally published in ref~\cite{EXP} and are reproduced by courtesy of the American Chemical Society (ACS). (b) Constant--height \gls{AFM} images of [19]dendriphene (first row), ball-and--stick depiction (right in the first row), interpolated images (framed by the grey dashed line) and the prediction performed by the \gls{CGAN} (second row). The \gls{AFM} images were published in ref~\cite{vilas201819} and are reproduced by courtesy of the European Chemical Societies Publishing.}
%\label{Fig:Pred_Experimental_Fakes}
%\end{figure}
%---------  FIG JMOL AFM ------------------- 

% dibenzothiophene 
The prediction of the model has not been so accurate in all experimental tests. \Cref{Fig:Exp_Test} (d and e) shows the test performed with \gls{AFM} images of dibenzothiophene and [19]Dendriphene respectively. In the dibenzothiophene prediction, the model gets right both the number of rings and the number of vertices in each ring, which is clear in the \gls{AFM} images taken at shorter tip--sample distances. % However, the translation performed by the model does not allow 
However, the model is not able to rescale the central ring to show the bonds with their correct size. Furthermore, although the model manages to reveal a slight yellow color at the sulphur apex, the size of the bonds in the prediction is larger than in the target, so the prediction is not conclusive. %Although the experimental \gls{AFM} images are taken on a tilted molecule on the adsorption surface. %On the one hand, it has not been possible to remove the noise from these images, and on the other hand, the deformation of the central ring is much more %severely 
%deformed in the experimental images than in the theoretical simulations. 
%Despite applying a filter to remove noise from the images, this objective was not achieved. 
It has to be noticed that, despite applying a filter, we were not able to remove the experimental noise completely. Furthermore, the central ring appears, for some unknown reason, much more deformed than in the theoretically simulated images. These two features of the experimental images may account for the failure of the prediction.
% \edR{However, this only reinforces the potential of the model.}
However, our previous work~\cite{Carracedo2021MDPI} shows that these problems with experimental images can be fixed. %\color{red} In ref.~\cite{Carracedo2021MDPI}, we demonstrated an \color{blue} 
We proposed a strategy that significantly improves the accuracy in the classification of a small set of molecules, including dibenzothiophene, from experimental images. We implemented a \gls{VAE} to incorporate, from just three experimental images, characteristic features into the training set that produce an increase in the accuracy of 0.28 (from 0.62 to 0.90) in the particular case of  dibenzothiophene and an increase of 0.2 for the whole set of molecules. This strategy can be extended to our \gls{CGAN} model to incorporate during the training images containing experimental features in order to improve its accuracy.

%. [19]dendriphene
The [19]dendriphene prediction is also partly a failure. Although it reveals a large part of the structure, it does not close five of the six peripheral rings. Moreover,  while in most cases, the prediction of the presence of carbon atoms is correct, the model tints some areas of the structure with bluish tones that do not allow to conclusively determine whether the chemical species is a carbon or a nitrogen. It has to be noticed that the test has been performed with only three experimental images, that is, less than a third of the information with which the model was trained. 
%It is an overly complicated test. With this input, at least the number of vertices of each revealed ring is correct. 
At the same time, it is also remarkable, that, even for such a complicated test and with a very limited input information, the number of vertices of each revealed ring is correct.

%In contrast to the test with theoretical images, 
%The experimental test has been performed with very few examples that neither provide sufficient images nor are in the range of tip--sample distances in which the \gls{AFM} images of \gls{AFMD} have been simulated. Despite these drawbacks, the results are promising. An experimental collaboration in which \gls{AFM} images are systematically taken in the conditions in which \gls{AFMD} was simulated, both for training the model and for testing, is necessary to assess the potential of the model.

%\edited{Conclusion: We need an experimental collaboration to fully adress the potential of the method. That is, experimental images under the conditions under which the model has been trained. So making it possible for something systematic to provide better results.}

\section{Discussion}

%1.-Classification, iupac to image to image translation

In summary, our results show the potential for chemical and structural identification of molecules encoded in \gls{AFM} images. %Similarly to the idea developed in \cref{Chap:Iupac}, 
%We propose a \gls{CGAN} to generalise the identification proposed in our previous works~\cite{Carracedo2021MDPI,Carracedo_IUPAC}. 
%%This approach goes beyond the identification method developed in  \cref{Chap:Iupac}, where we were forced to limit the number of terms included in the \gls{IUPAC} name of the molecule.
We propose a \gls{CGAN} to generalise the accurate classification of a small set of molecules achieved in our previous work~\cite{Carracedo2021MDPI} into a general purpose tool to completely determine the structure and composition of arbitrary organic molecules. Our model performs a direct translation between a stack of 10 constant--height \gls{AFM} images and the ball--and--stick depiction of the molecule. We are only limited by the fact that the atoms composing the molecule have to be in the training dataset. Since \gls{AFMD}~\cite{QUAM-AFM_repository} includes the most relevant chemical species in organic chemistry, the model prediction is practically unconstrained.

%Moreover, this approach also has slight improvements over the classification performed with \gls{RNN} for the prediction of the \gls{IUPAC} name. In the case of \gls{IUPAC}, it can predict any combination of the terms shown in \cref{Subsec:IUPAC_DATA}, however, there are some terms that have been discarded because they are not sufficiently repeated in the dataset. Even some structures were discarded due to unusual symbols such as superscripts. In this chapter we performed a translation between the \gls{AFM} image stack and the ball--and--stick depiction. Thus, the only dependence of the identification is that the atoms composing the molecule are in the dataset. Since \gls{AFMD} is composed of the chemical species that define organic chemistry, the model prediction is unconstrained.

%4.- Good accuracy in simulations and experimental images with different configurations because the structures are not flat.

Molecular identification in both theoretical and experimental images is highly accurate with a model trained exclusively with theoretical images. The ability of the model to reveal molecular structures and chemical species is truly remarkable, beyond the capabilities of a human expert in the field. Moreover, these identifications are not conditioned to a single molecular configuration, since the differences in height of the atoms in the gas-phase structures included in the training dataset provide enough information to identify patches of the image according to the chemical environment of each atom. In this way, the model has learned to decipher the distortions produced by each chemical species in relation to its surroundings regardless of the relative height difference in the molecule.

%2.- The model generalises the identification in a similar way to the model for IUPAC prediction but has a better accuracy due to the consistency of CNNs instead of RNNs.

%\gls{CGAN} predictions for identification have better accuracy than those developed previously \cite{Carracedo2021MDPI,alldritt2020automated,Carracedo_IUPAC}. Firstly, 
We attribute the high performance of the model to the consistency and robustness shown by  \gls{CNN}s in the analysis of images with \gls{DL}, together with the patch analysis performed by the discriminator and the use of a suitable loss function, with an L1 distance, that increase the sharpness of the predictions and makes the mapping between input and output accurate. The reduced accuracy shown for structures that have a very high internal torsion is not a critical issue when facing the identification from experimental \gls{AFM} images, as real adsorbed structures tend to be flatter than the corresponding gas--phase ones. Moreover, in these high-torsion cases, the model correctly reveals both the structure and the chemical species located on the top areas of the molecule. The presence of atoms in the lower areas is indicated with bonds that are eventually blurred due to the lack of information. Thus, more than a problem of the model, this reduced accuracy represents  an intrinsic limitation of the current \gls{AFM} set-ups, that may be fixed by an alternative operation mode. 

The few results presented for molecular identification based on experimental AFM images, in spite of the incomplete information available, are really promising.  
An experimental collaboration in which \gls{AFM} images are systematically taken in the conditions in which \gls{AFMD} was simulated, both for training the model and for testing, would be necessary to properly assess the potential of the model.

\section{Methods}

%\subsection{QUAM-AFM Images}
\subsection{QUAM-AFM Data Set}

\gls{DL} models need large datasets to adjust the weights in each of their layers. In this work, we take advantage of \gls{AFMD}~\cite{QUAM-AFM_repository}, an open-access dataset that includes simulations of theoretical AFM images, based on the latest HR-AFM modeling approaches~\cite{ellner2019molecular,liebig2020quantifying,QUAM-AFM_repository}, for a collection of 686,000 molecules that include 10 different atomic species (C, H, N, P, O, S, F, Cl, Br, I). Here we provide the main characteristics that are relevant for our study and refer the reader to the original publication~\cite{QUAM-AFM_repository} for details.
%  Restriction to quasi-planar molecules.
\gls{AFMD} focuses on quasi-planar molecules, that is, molecules which display height variations up to 1.83~\r{A} along the $z$--axis 
%This distance is larger than 1.5~\AA\ --the descriptor used in ref.~\cite{alldritt2020automated} as the height range where estructural information can be retrieved from 3D structures with a collection of \gls{AFM} images taken at different heights-- 
in order to include  aliphatic chains and $sp^{3}$ carbon atoms (methyl groups) as possible side groups.

% operational parameters
The contrast of \gls{AFM} images taken in the  FM mode with CO--metal tips depends on parameters, such as the cantilever oscillation amplitude or the average tip-sample distance,  that can be controlled during operation, and also on the tip nature, in particular, differences in the attachment of the CO molecule to the metal tip that have been consistently observed and characterised in experiments~\cite{liebig2020quantifying,weymouth2014quantifying,NeuPRB2014}. In order to cover the widest range of variants in the \gls{AFM} images, \gls{AFMD}  was simulated with 6 different oscillation amplitudes of the cantilever (0.40, 0.60, 0.80, 1.00, 1.20, 1.40 ~\AA), 10 tip-sample distances (2.80, 2.90, 3.00, 3.10, 3.20, 3.30, 3.40, 3.50, 3.60, 3.70 ~\AA),  an 4 values of the elastic constant describing the tilting stiffness of the CO-metal bond (0.40, 0.60, 0.80, 1.00 N/m). These 240 combinations are applied to each of the molecular structures, resulting in a total of 165 million grey-scale images with resolution $256\times256$ pixels. 
 \gls{AFMD} also provides the ball-and-stick depictions of each molecule generated from the atomic coordinates. 
%  and therefore, proportionally to the \gls{AFM} images. That is, 
These depictions share the same scale used in the \gls{AFM} images: if we superimpose the two images, each ball of the representation is centered on the position occupied by the atom it represents in the \gls{AFM} images. 

\subsection{\gls{CGAN} Molecular identification model} \label{Sec:CGAN_Details}

The generator for the identification of molecules through \gls{AFM} images is composed of a series of similar blocks where the main difference is the number of kernels applied in each convolution and the dimensions of each input (see~\cref{Fig:CAN}(a)). The input consists of a stack of 10 greyscale \gls{AFM} images (a single channel).  This stack  is processed in a dropout layer, with a rate of 0.5, followed by two 3D convolutional layers. The first 3D convolution includes 64 kernels, each of them has (4,3,3) size and is applied with a stride of (3,1,1) and padding. The second 3D convolution also has 64 kernels but, in this case, the kernels have  size (4,4,4) and are applied with a stride of (4,2,2). The output of the second convolutional layer is resized to (128,128,64) and activated with a \gls{lrelu} function. 

From this point on, the encoder consists of seven blocks, represented by yellow boxes in  \cref{Fig:CAN}(a). Each block includes %consisting of a sequence of 2D convolution, 
a 2D convolution followed by a batch normalisation and a \gls{lrelu} activation function with $\alpha=0.2$. All kernels of the 2D convolution have size (4,4) and are applied with a stride of (2,2). The 2D convolutional layers have 128, 256, 512, 512, 512, 512,  and 512 kernels,  taking as reference the processing direction from the one closest to the input to the one closest to the compressed representation space. The outputs of the activations are used both to feed the next block of the encoder and to feed the decoder block of the same size. The generator decoder blocks, represented by green boxes in \cref{Fig:CAN}(a), include the following layers: a  transposed convolution, a batch normalization, a dropout layer with rate 0.2 (only in the three layers closest to the space of the compressed representation, see \cref{Fig:CAN}), a concatenation with the output of the corresponding encoder block, and, finally, a \gls{relu} activation (except for the last block, the one closest to the output,  that is activated with an hyperbolic tangent function).
%\edR{Methods??:Once the stacks of 10 tip--sample distance images were generated, we denoised them by applying the \textit{medianBlur} filter with size 3 from the OpenCV Python package to all input and output images.} 
%\edR{Finally a \textit{medianBlur} filter with size 3 from the OpenCV Python package is applied to the output images.}

The discriminator (\cref{Fig:CAN}(b)) consists of a sequence of layers, initiated by a concatenation of all input images (note that we can consider the 10 \gls{AFM} images as a single image with 10 channels). It is followed by a 2D convolutional layer with 64 kernels of size (4,4) and stride of (2,2) activated with \gls{lrelu}. Then, it has four blocks consisting of a 2D convolutional layer, a batch normalization and a \gls{lrelu} activation ($\alpha=0.2$). The convolutions have 128, 256, 512 and 512 kernels with size (4,4) and stride (2,2) respectively. The last layer is a 2D convolution with a single kernel of size (4,4) which is activated with the sigmoid function.

\subsection{\gls{CGAN} Training} \label{Sec:CGAN_Training}

% Training
The 686K structures in \gls{AFMD} have been split into training, validation and test sets with 581K, 24K and 81K structures respectively. % complying with the best practices in machine learning~\cite{artrith2021best}.  
%In this way, \gls{AFMD} complies with all the requirements set out in ref. \cite{artrith2021best}. 
The test set is chosen to be particularly large for two reasons. Firstly, to perform a quantitative analysis with randomly chosen structures in order to avoid an statistical fluke. Secondly, it is desirable to have sufficient variety of structures to be able to show examples that reflect the most salient strengths and weaknesses of the model.

%During training, we use a random selection of the \gls{AFM} simulation parameters for each molecule in the training set at each epoch. This variability in the input data means that the parameters with which the \gls{AFM} experiment has been carried out are not decisive for the model to succeed or fail in recognising the structure. On the other hand, techniques commonly applied in \gls{DL}, consisting of deforming the input images so that the model learns different features, have been applied (see \cref{Fig:IDG}).
% %we have applied techniques commonly applied in \gls{DL} that consist of deforming the input images so that the model learns different features (see \cref{Fig:IDG}). 

%\edR{Original: During training, we use a random selection of the \gls{AFM} simulation parameters for each molecule in the training set at each epoch.} 
During training, we randomly choose one of the combinations of \gls{AFM} simulation parameters available in \gls{AFMD} for each input stack. % \color{blue} NO! We use a combination of parameters for each set of input images, in this new text we say that it is the same combination for the whole epoch and this is not true. \color{black}
This variability in the input data makes sure that the parameters with which the \gls{AFM} experiment has been carried out do not play a decisive role in the success of the identification, prevents overfitting,  and provides the model with the ability to generalize. 
% IDG
 This variability is further enhanced with the application of an \gls{IDG} to the training  set.  This technique, commonly used in \gls{DL}, applies different deformations (zoom, rotations, shifts, flips and shear) to the input images.   Let's recall that the ball--and--stick depictions included in \gls{AFMD} share the same scale as the \gls{AFM} images.
Thus the \gls{IDG} has to be applied to both the input \gls{AFM} images and the ball--and--stick depiction during the training: i. e.,  if we rotate the input \gls{AFM} images, then, the corresponding ball--and--stick depiction must be rotated with the same angle. Otherwise the atomic positions of the ball--and--stick representation would not match the corresponding atomic positions of the \gls{AFM} images, and the \gls{CGAN} would not be able to learn a local translation (from the pixel environment) between the shape and intensity of the \gls{AFM} image and the type of atom that caused it. This applies to all the operations in the \gls{IDG} except for the shear, that is not applied to the output ball--and--stick depiction. This is motivated by the fact that shear represents  a deformation that may appear in the experiments due to noise or tip asymmetries but it should not be present in the prediction. 

% IDG parameters matter !
We have found that the selection of appropriate deformation parameters for the \gls{IDG} applied to the training set during the fitting considerably increases the accuracy of the model in the test carried out with experimental images~\cite{Carracedo2021MDPI}. An particular example of the application of the \gls{IDG} and information on the range values used for the different operations can be found in Fig.~S3.

% Details for training 
Regarding the loss functions, the generator of the \gls{CGAN} was compiled with \gls{MAE} (using the parameter $\lambda=100$ defined by Isola~\textit{et al.}~\cite{isola2017image}) ,  while the binary cross entropy was used for the discriminator. The model was minimised by applying batches of 32 inputs with the \gls{Adam} optimiser, where the learning rate and first moment parameters were set to $2\cdot 10^{-4}$ and $0.5$ respectively. The training of the model was carried out during six epochs (109K iterations), displaying 300 predictions of the validation set to estimate the optimal training point every 10.000 iterations.

\section{Acknowledgments}
We would like to acknowledge support from the Comunidad de Madrid Industrial Doctorate Programme 2017 under reference number IND2017/IND-7793 and from Quasar Science Resources S.L. R.P. acknowledge support from the Spanish Ministry of Science and Innovation, through project PID2020-115864RB-I00  and the ``Mar\'{\i}a de Maeztu'' Programme for Units of Excellence in R\&D (CEX2018-000805-M). Computer time provided by the \gls{RES} at the Finisterrae II Supercomputer is also acknowledged.

\newpage

\appendix

\newpage

\bibliographystyle{Science}
%\bibliography{DL_AFM}
\bibliography{DL_AFM_IUPAC}
\addcontentsline{toc}{chapter}{Bibliography}

\end{document}


\baselineskip24pt

\maketitle

\newpage

\section{Model Accuracy with Simulated AFM Images}

\begin{table}[b!]
\centering
\begin{tabular}{| l| c| c | c | c |}
\hline
\hline
z diff.& Support& Acc.Tot.& Struct.Acc.& Acc. Atoms  \\ 
\hline
\hline
[0,183] pm & 3015 & 0.74 & 0.95 & 0.96 \\  
\hline
[0,50) pm & 2301 & 0.84 & 0.98 & 0.98 \\  
\hline
[50,100) pm & 294 & 0.56 & 0.90 & 0.92 \\  
\hline
[100,150) pm & 378 & 0.32 & 0.82 & 0.86 \\  
\hline
[150,183] pm & 42 & 0.20 & 0.60 & 0.82 \\  
\hline   
\end{tabular}
\caption{Accuracy of the model in a test where both the 3015 structures and their simulation parameters have been randomly selected.  The table shows (from left to right) the range of maximum height differences between the atoms of each molecular structure in the evaluated ensemble, the number of structures in the evaluated set, the overall accuracy (perfect structure and atom prediction), the accuracy of structure discovery, and the accuracy in revealing the atomic species. The accuracy of the structures has been evaluated considering that the final result is correct only if  the prediction is perfect: it shows all  of the bonds of the molecule and the number of vertices in each structure (chain or rings), with the exception of the hydrogens and its bonds. The structure accuracy has been calculated as the percentage of fully discovered (perfect) structures out of the total set of structures. % so that the number of wrong bonds in the incorrect predictions has not been weighed. 
The accuracy in the prediction of the atomic species has been evaluated as the percentage of total hits (correct predictions) over the total number of atoms in the set, 
without considering the hydrogens.} %and the accuracy value shown is the percentage of total hits over the number of atoms in the set.}
\label{Table:Accuracy_Th}
%\end{table}
\end{table}

The objective of our \gls{CGAN} is the identification of molecules through experimental \gls{AFM} imaging. However, a test with a set of computationally simulated images sheds light on the actual capability of the model due to the control we have over the molecular features. In particular, we emphasize the strong dependence of the model on the difference in atom heights. To assess this dependence it is required to have absolute control over the molecular torsion, which is not possible with current experimental techniques. \Cref{Table:Accuracy_Th} shows the details of a test set of 3015 molecules. For this purpose the test set has been divided into four subsets according to the maximum difference in height of the atoms in the molecule (considering the height as the distance as the distance between atoms measured perpendicularly to the molecular plane) in the performance of the model.  We emphasize that the structures used in the simulation are in gas phase, which leads some of them to have a strong torsion. In case of depositing these molecules on a surface, the internal torsion would decrease significantly due to the interaction with the substrate, and almost all of them would belong to the range of less than 50 pm, where the result of the simulation would be closer to the images obtained in the experiments that are necessarily carried on molecules adsorbed on a substrate.

\Cref{Table:Accuracy_Th} shows the power of the model to resolve molecules with very high accuracy in cases where the molecule is in a roughly flat configuration. Moreover, there seems to be a limiting distance beyond which the model cannot provide information. The motivation for this decrease in accuracy lies in the constant height technique used to obtain the images, where the distance between the tip and the atoms closest to the surface is too long for the tip-sample interaction to provide sufficient information for identification.
As discussed in the main text, in spite of these limitations, the training with gas--phase structures that have a high internal torsion, rather than being a limitation, is enhancing the ability of the model to generalize and to recognise molecules in different adsorption configurations.

\section{Experimental Test Results}

\begin{figure}[p!]
\centering
\includegraphics[clip=true, width=1.0\columnwidth]{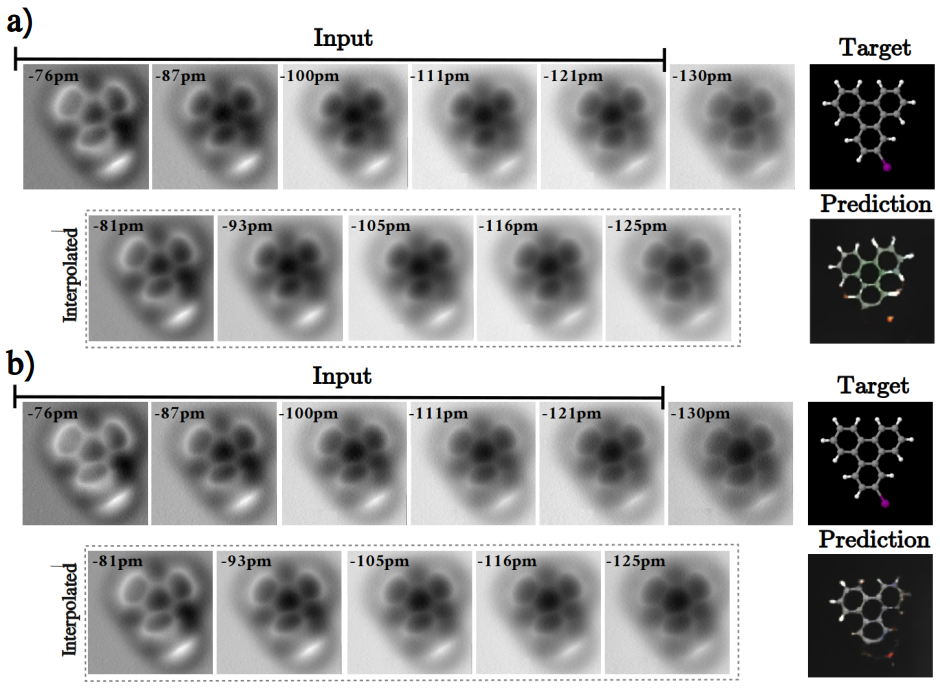}
 \caption{(a) phase modulation \gls{AFM} images and (b) frequency shift (FM) images for 2-iodotriphenylene on Ag(111). First row of each panel shows the six experimental \gls{AFM} images originally published in Figure S4 in ref.~\cite{martin2020bond} and reproduced by courtesy of the American Physical Society (APS). The second row shows the results of the interpolation. The right--hand column shows the ball--and--stick depiction of the structures and the predictions performed by the model. 
In spite of the different operation modes (FM images are taken with different oscillation amplitudes) and the reduced tip--height range of 64pm (compared to100 pm in training), our \gls{CGAN}  is able to fully reveal the ring structure and the presence of halogen atom although the predicted color is more like the one associated to bromine than the one  corresponding to iodine.
Phase images where taken in constant height scans while operating the AFM in amplitude modulation with Q--control (Q$_{eff}=2060$) using  free oscillation amplitudes of 45, 70, 114, 175, 280, and 525 pm.  FM images where also taken constant height scans using the same oscillation amplitudes. Before each phase image, the STM feedback was activated above the Ag(111) surface with the same tunneling parameters (I =10 pA, U = 7 mV) and the desired oscillation amplitude in FM mode. The average distances between the tip and the substrate were then chosen with respect to this reference so that similar contrast features were shown in the phase and FM images and also in the amplitude images shown in Fig. 5 in the main text. These average tip-substrate distances are 76, 87, 100, 111, 121, and 130 pm for the different oscillation amplitudes, as marked in the figure.}
\label{Fig:Pred_Experimental_Iodothiph}
\end{figure}

The scarcity of \gls{AFM} images of the same molecule at different tip-sample distances available in the literature has led us to test the model with all the images found. In most of these tests the conditions and techniques used are far from those used to simulate \gls{AFMD}~\cite{QUAM-AFM_repository}, which has been the data set used to train the model and therefore defines the conditions under which the images should be obtained experimentally to perform the test. \gls{AFMD} uses various oscillation amplitude ranges for the simulations, as well as tilt stiffness constants for the CO molecule. These parameters have been used during the training of the network to enable it with a greater capacity for generalization. However, 10 input images covering a range of 100 pm tip-to-sample distance magnification (280-380 pm tip-to-sample distances) have always been used. Some examples of tests carried out with images obtained from different configurations have been shown in the main text. Here we add some more examples where, although the predictions are not perfect, they reinforce the potential of the model.

\Cref{Fig:Pred_Experimental_Iodothiph} shows six phase (top of panel a) and frequency modulation (FM) (top of panel b) \gls{AFM}  images published in Figure S4 in the supplementary material of ref.~\cite{martin2020bond}, together with the interpolations that we have made to complete the 10--image stack needed as input for our CGAN, and the predictions of our \gls{CGAN}. Phase images where taken in constant height scans while operating the AFM in amplitude modulation with Q--control (Q$_{eff}=2060$) using  free oscillation amplitudes of 45, 70, 114, 175, 280, and 525 pm.  FM images where also taken constant height scans using the same oscillation amplitudes. Before each phase image, the STM feedback was activated above the Ag(111) surface with the same tunneling parameters (I =10 pA, U = 7 mV) and the desired oscillation amplitude in FM mode. The average distances between the tip and the substrate were then chosen with respect to this reference so that similar contrast features were shown in the phase and FM images and also in the amplitude images shown in Fig. 5 in the main text. These average tip-substrate distances are 76, 87, 100, 111, 121, and 130 pm for the different oscillation amplitudes, as marked in the figure.

Before looking at the predictions of our CGAN, it has to be stressed that none of these AFM images correspond to the AFM operation mode used to simulate the AFM images employed in the training of the model. This is clear for the phase images but it is also the case in the FM images, as the oscillation amplitude is different in each of the images while the10--image stacks used in the training correspond to different tip--sample distances of closest approach but to the same oscillation amplitude. Moreover,  the tip--height  range covered by the images (64 pm) is significantly smaller than the 100pm that we consider optimal. 
In spite of these strong limitations in the input, our \gls{CGAN}  is able to fully reveal the ring structure and the presence of the halogen atom from the phase images shown in \Cref{Fig:Pred_Experimental_Iodothiph}(a),  although the model is not able to have a clear prediction about the halogen, since the color is more like the one associated to bromine than the one  corresponding to iodine. A similar result is shown in \cref{Fig:Pred_Experimental_Iodothiph}(b) for the FM images. In this case, the model predicts a very weak ring halo around the halogen that should not be present. Far from considering these predictions a failure, these results indicate that our CGAN model can provide very useful information regarding the molecular  identification when fed with images taken with different AFM operation modes, provided that the atoms of these molecules have been included in the training set. %provided that these atoms have been included in the training set.

\begin{figure}[t!]
\centering
\includegraphics[clip=true, width=1.0\columnwidth]{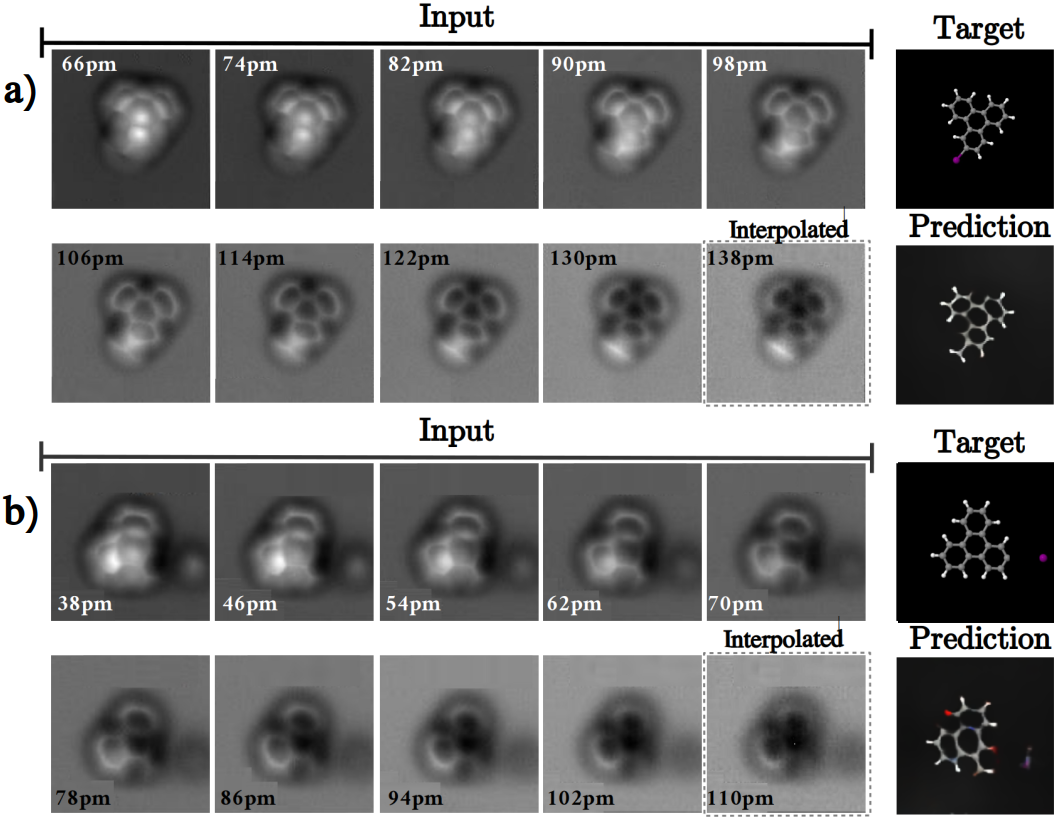}
 \caption{Constant-height AFM images taken in Frequency Modulation (FM) mode for (a) 2-iodotriphenylene on Ag(111) and for (b) the products of a dehalogenation reaction locally triggered using a voltage:  a triphenylene (TP) radical and the cleaved I atom.  \gls{AFM} images were originally published in Figs~S1(g) and S2(g) in the supplementary material of ref.~\cite{martin2019bond} and are reproduced by courtesy of the American Physical Society (APS). 
The image features at the halogen position and its evolution with tip height in (a) are quite different from those shown in other experimental examples and from our AFM simulations, and the model predicts a methyl group instead of a halogen.  In (b), our model captures the presence of the cleaved I atom and provides a strongly deformed structure where the dehalogenated ring is not closed, consistent with the lack of information in the AFM images due to the strong bending of the molecule towards the substrate induced by the interaction of the unsaturated C bond in that ring with the metal.} 
\label{Fig:Pred_Experimental_Iodothip_2h}
\end{figure}

\Cref{Fig:Pred_Experimental_Iodothip_2h} shows the predictions of our CGAN made with the images published in the supplementary material of ref.~\cite{martin2019bond} 
\Cref{Fig:Pred_Experimental_Iodothip_2h}(a) corresponds to a series of frequency shift images of 2-iodotriphenylene in constant height-mode (Vgap = -0.57 mV, A = 52 pm) taken from Fig~S1(g) in ref.~\cite{martin2019bond} and the corresponding prediction. The distance values (z values) are given with respect to a tunneling current of I = 10 pA and a gap voltage of Vgap = 4 mV above the Ag(111) surface (see ref.~\cite{martin2019bond} for details). In this case, the model fails to predict the presence of the halogen, which is replaced by a methyl group. A priori, this should be a simple case, since there are 9 \gls{AFM} images and the tip-to-sample distance increments are similar to those used to simulate the \gls{AFMD} image stacks. %\color{red} However, the image features at the halogen position are quite different from those shown in other examples, and this makes the model unable to identify the iodine. \color{black}
However, for some unknown reason, the image features at the halogen position and its evolution with tip height are quite different from those shown in other experimental examples and from our AFM simulations. We speculate that this is the reason behind the failure of the model in the identification of the iodine. More work on both the experimental and theoretical side would be required in the future to properly understand this case.
%\edR{RP: confunde halogenos y sp3 en las imagenes teoricas??, tenemos simulaciones de esta molecula con esos dos grupos terminales??}
% 2-Iodotriphenylene | C18H11I | CID 88955426
% 2-Methyltriphenylene | C19H14 | CID 618051 

Lastly, \cref{Fig:Pred_Experimental_Iodothip_2h}(b) shows a series of frequency shift images in constant-height mode (Vgap = -0.57 mV, A = 52 pm) of the products of a dehalogenation reaction locally triggered using a voltage:  a triphenylene (TP) radical and the cleaved I atom. (Fig~S2(g) in ref.~\cite{martin2019bond}) The distance values (z values) are given with respect to a tunneling current of I = 10 pA and a gap voltage of Vgap = 4 mV above the Ag(111) surface. As discussed in detail in ref.~\cite{martin2019bond}, the dehalogenated aryl ring of the TP radical bends strongly towards the surface.  Consequence of this effect, the interaction of the ring with the CO tip is very weak, which causes it not to be visible in the \gls{AFM} images, enhancing the black halo in this area. Our model predicts a strongly deformed structure --where the ring is not closed and chemical species like O and H are saturating the bonds-- and the presence of an isolated I atom, not bonded to the molecule. The ability to detect of the presence of the halogen atom is really remarkable while the failure in showing the strongly bent aryl ring reflects more a limitation of the AFM operation mode (ref.~\cite{martin2019bond} proposes the use of a novel constant current mode to gather information on these highly corrugated cases) than an intrinsic failure of our model.

%\edR{(a) from Fig~S1(g) in ref.~\cite{martin2019bond}: Series of frequency shift images of ITP in constant-height mode (Vgap = -0.57 mV, A = 52 pm). The distance values (z values) are given with respect to a tunneling current of I = 10 pA and a gap voltage of Vgap = 4 mV above the Ag(111) surface [see gray point in Fig. S1(b)]. The first and the last images of this series are depicted in Figs. 2(a)-(f).}

%\edR{(b) from Fig~S2(g) Series of frequency shift images of TP radical !!! in constant-height mode (Vgap = -0.57 mV, A = 52 pm). The distance values (z values) are given with respect to a tunneling current of I = 10 pA and a gap voltage of Vgap = 4 mV above the Ag(111) surface. The dehalogenated aryl ring of the TP radical is bent towards the surface (Fig. 4).!!!!}

\newpage

\section{Data Augmentation} \label{Sec:IDG_CGAN}

\begin{figure}[b!]
\centering
\includegraphics[clip=true, width=1.0\columnwidth]{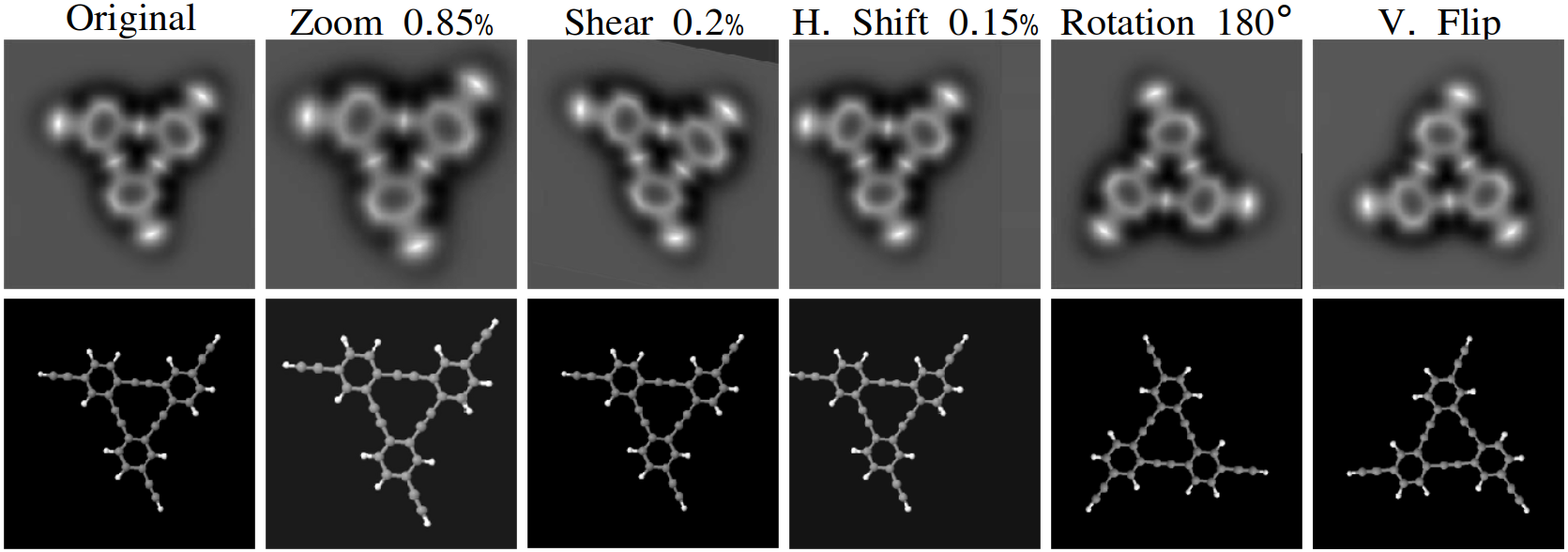}
 \caption{\gls{IDG} applied to \gls{AFM} input images and ball--and--stick depiction for the molecule 6,14,22-triethynyltetracyclo[18.4.0.04,9.012,17]tetracosa-1(20),4(9),5,7,12(17),13,15,21,23-nonaen-2,10,18-triyne.  The parameters selected for the \gls{IDG} during the training are randomly chosen in the range of [-180,180] degrees for rotation, [-15,15]\% for zoom and for vertical and horizontal shift, [-20,20]\% for shear and random vertical and horizontal flip. When necessary, nearest filling has been applied for the points outside of the boundaries of the input.}% \edR{A proper selection of parameters for the \gls{IDG}  considerably increases the accuracy of the model in the tests carried out with experimental images.}}
\label{Fig:IDG}
\end{figure}
%---------  FIG JMOL AFM -------------------

As we discussed in the main text, both the random selection of the simulation parameters of each input stack and the deformation of the images are key to achieve identification in the experimental images. The technique of deforming the input images is known as data augmentation and is commonly used in computer vision. This makes the model capable of recognizing molecules through more complex images than those in the training set. 

The values selected for these distortions in this work are randomly chosen in ranges of [-180,180]--degree rotations, $\pm0.15$ zoom range,  $\pm0.2$ shear range, and $\pm0.15$ both vertical and horizontal shift range, as illustrated in \cref{Fig:IDG}. When a molecule is rotated or moved during an \gls{AFM} experiment, all resulting images show similar variations, so we apply the same deformation parameters to the ten images that compose each stack. As discussed in the main text, the \gls{IDG} should be applied to both the input \gls{AFM} images and the ball and stick representations during training (with the exception of shear, which is not applied to the output ball and stick representation).

%\afterpage{\blankpage}

%\appendixstart
\appendix
\renewcommand\thefigure{\thesection.\arabic{figure}}    

\setcounter{figure}{0}   

%\section{Appendix}

\newpage

\bibliographystyle{Science}
%\bibliography{DL_AFM}
\bibliography{supplemental}
\addcontentsline{toc}{chapter}{Bibliography}